\documentclass{aa}  

\usepackage{graphicx}
\usepackage{natbib,graphicx}
\bibpunct{(}{)}{;}{a}{}{,} % to follow the A&A style
\newcommand{\kms}{\,km\,s$^{-1}$}

\begin{document}

\title{The Submillimeter Array 1.3\,mm line survey of Arp\,220}


\author{
  S. Mart\'in\inst{\ref{inst1}}
  \and M. Krips \inst{\ref{inst2}}
  \and J. Mart\'in-Pintado\inst{\ref{inst3}}
  \and S. Aalto\inst{\ref{inst4}}
  \and J.-H. Zhao\inst{\ref{inst5}}
  \and A.B. Peck\inst{\ref{inst6}}
  \and G.R. Petitpas\inst{\ref{inst7}}
%  \and T. Greve\inst{\ref{inst8}}
  \and R. Monje\inst{\ref{inst8}}
  \and T.R. Greve\inst{\ref{inst9}}
  \and T. An\inst{\ref{inst10}}
}
\institute{
  European Southern Observatory, Alonso de C\'ordova 3107, Vitacura, Casilla 19001, Santiago 19, Chile\\
  \email{smartin@eso.org}\label{inst1}
  \and
  Institut de Radioastronomie Milim\'etrique, 300 rue de la Piscine, 38406 Saint Martin d'Heres, France
  \label{inst2}
  \and
  Centro de Astrobiolog\'ia (CSIC-INTA), Ctra. de Torrej\'on Ajalvir, km. 4, E-28850 Torrej\'on de Ardoz, Madrid, Spain
  \label{inst3}
  \and
  Department of Earth and Space Sciences, Chalmers University of Technology, Onsala Observatory, SE-439 92, Onsala, Sweden
  \label{inst4}
  \and 
  Harvard-Smithsonian Center for Astrophysics, 60 Garden Street, Cambridge, MA 02138, USA
  \label{inst5}
  \and
  Joint ALMA Observatory, Alonso de C\'ordova 3107, Vitacura, Casilla 19001, Santiago 19, Chile
  \label{inst6}
  \and
  Harvard-Smithsonian Center for Astrophysics, Submillimeter Array, 645 North A‘ohoku Place, Hilo, HI 96720, USA
  \label{inst7}
  \and
  California Institute of Technology, Cahill Center for Astronomy and Astrophysics 301-17, Pasadena, CA 91125, USA
  \label{inst8}
  \and
  Dark Cosmology Centre, Niels Bohr Institute, University of Copenhagen, Juliane Maries Vej 30, DK-2100 Copenhagen \O, Denmark
  \label{inst9}
  \and
  Shanghai Astronomical Observatory, Chinese Academy of Sciences, Shanghai 200030, China
  \label{inst10}
}

\abstract
{
Though Arp\,220 is the closest and by far the most studied ULIRG, a discussion is still ongoing on the main power source
driving its huge infrared luminosity.
}
{
To study the molecular composition of Arp\,220 in order to find chemical fingerprints associated with the main heating
mechanisms within its nuclear region.
}
%leading to the observed chemistry.}
{
We present the first aperture synthesis unbiased spectral line survey toward an extragalactic object.
The survey covered the 40~GHz frequency range between 202 and 242~GHz of the 1.3~mm atmospheric window.
}
{
We find that 80\% of the observed band shows molecular emission, with 73 features identified from 15 molecular species and
6 isotopologues. The $^{13}$C isotopic substitutions of HC$_3$N and transitions from H$_2^{18}$O, $^{29}$SiO, and CH$_2$CO
are detected for the first time outside the Galaxy.
No hydrogen recombination lines have been detected in the 40~GHz window covered.
The emission feature at the transition frequency of H31$\alpha$ line is
identified to be an HC$_3$N molecular line, challenging the previous detections reported
at this frequency.
%No hydrogen recombination lines are detected, challenging previous detections reported at these frequencies.
Within the broad observed band, we estimate that 28\% of the total measured flux is due to the molecular line contribution,
with CO only contributing 9\% to the overall flux.
We present maps of the CO emission at a resolution of $2.9''\times1.9''$ which, though not enough to
resolve the two nuclei, recover all the single-dish flux.
The 40~GHz spectral scan has been modelled assuming LTE conditions and abundances are derived for all identified species.
}
{
The chemical composition of Arp\,220 shows no clear evidence of an AGN impact on the molecular emission
but seems indicative of a purely starburst-heated ISM.
The overabundance of H$_2$S and
the low isotopic ratios observed suggest a chemically enriched environment by consecutive bursts of star formation, with
an ongoing burst at an early evolutionary stage.
The large abundance of water ($\sim10^{-5}$), derived from the isotopologue H$^{18}_2$O, as well as the
vibrationally excited emission from HC$_3$N and CH$_3$CN are claimed to be evidence of massive star forming regions within Arp\,220.
Moreover, the observations put strong constraints on the compactness of the starburst event in Arp\,220.
We estimate that such emission would require $\sim2-8\times10^6$ hot cores, similar to those found in the Sgr~B2 region in the Galactic
center, concentrated within the central 700~pc of Arp\,220.
}



\maketitle



\section{Introduction}
At a redshift of $z=0.018$, Arp\,220 is the closest ultraluminous infrared galaxy (ULIRG).
This galaxy is  an advanced merger system as evidenced by the large tidal tails observed in the optical \citep{Joseph1985,Kim2002,Koda2009}
and the double nuclei (separation $\sim0.98''$)
%separated by only $0.98''$,
observed in radio \citep{Norris1988}, mm \citep{Scoville1997,Downes1998,Sakamoto1999}, sub-mm \citep{Sakamoto2008} and near-IR wavelenghts \citep{Graham1990,Scoville1998}.
These nuclei are surrounded by two counterrotating gas disks as well as a larger outer disk encompassing both \citep{Sakamoto1999,Mundell2001}.
The main nuclear powering sources in ULIRGs are thought to be starbursts events, active galactic nuclei (AGN), or a combination of both.
Mid-IR surveys of ULIRGs with ISO \citep{Genzel1998} suggested that $70\%-80\%$ of the population are powered by star formation while only $20\%-30\%$ are AGN powered.
This result is consistent with that from near-IR searches for hidden broad-line regions (BLRs) towards ULIRGs that lack BLR signatures in
the optical \citep{Veilleux1999}.

The identification of the main power source becomes very elusive in extremely obscured nuclei like those of Arp\,220.
The nuclei in Arp\,220 are affected by a severe obscuration at $2.2\mu$m \citep{Scoville1998}.
Even at 1~mm, the dust towards the more massive western nucleus is found to be significantly
optically thick \citep[$\tau\sim1$,][]{Downes2007}.
Thus, a number of arguments have been proposed in favor of both powering scenarios
%as leading 
to explain the large observed IR luminosity of Arp\,220, though
none of them have been conclusive.

Among the evidence favoring the AGN-powered scenario are studies showing
hard X-ray emission strongly concentrated towards the nuclei, a hard spectrum X-ray point source close to the position of the Western nucleus,
and a softer point source towards the Eastern nucleus \citep{Clements2002}.
Additionally, the luminosity ratio $L_{X(2-10\,\rm keV)}/L_{FIR}$ is unusually low relative to what is observed in starburst galaxies \citep{Iwasawa2001,Iwasawa2005},
and the 1.98\,keV equivalent width of the 6.7\,keV line of Fe is too large to be purely starburst-driven \citep{Teng2009}.
The high column densities and therefore obscuration could hide a Compton thick AGN \citep{Sakamoto1999,Sakamoto2008,Downes2007}.
%High resolution HI observations \citep{Mundell2001} find a larger column density towards the eastern nucleus.

However, the starburst-driven scenario also appears to be supported by a number of observations.
The detection of dozens of radio supernovae (RSNe) in both nuclei of Arp\,220 is a clear indication of the starburst events in this galaxy \citep{Lonsdale2006,Parra2007}.
The supernova rate of $4\pm2\,\rm yr^{-1}$ agrees with the derived star formation rate based on its FIR luminosity \citep[$\sim300~M_\odot\,{\rm yr}^{-1}$,][]{Dopita2005}, which implies that the starburst
traced by the detected RSNe is able to produce the observed FIR luminosity \citep{Lonsdale2006}.
However this conclusion is based on assumptions about the truncation of the Initial Mass Function (IMF) and that all supernova events result in RSNe, which requires
a dense and compact starburst environment \citep{Smith1998}.
Towards the Eastern nucleus, the agreement between the supernovae and the dust surface density suggests a starburst heated dust \citep{Sakamoto2008}.
A comparison of the OH megamaser emission \citep{Lonsdale1998} with the sub-mm continuum peak suggests that even if the masers might be associated
to an AGN, it would not be the main contributor to the dust heating \citep{Sakamoto2008}.
Bright water vapor megamaser emission at 183\,GHz suggests the presence of $\sim10^6$ Sgr\,B2-like hot cores within the central kiloparsec of
Arp\,220 \citep{Cernicharo2006}.
Extended faint soft X-ray emission is detected beyond the optical boundaries of the galaxy with bright plumes extending 11\,kpc,
% across the nucleus,
claimed to be the result of the starburst driven superwinds \citep{McDowell2003}.

The dense gas phase of the interstellar medium (ISM) in Arp220 has been targeted by numerous observations
of molecular tracers emission in the mm and sub-mm wavelenghts.
%Within the context of mm and sub-mm molecular emission, Arp\,220 has been the target of numerous observations of emission from dense molecular tracers.
These observations provided additional constraints on the nature of
%have served to additionally constrain 
the power source from a ISM chemical composition point of view.
HNC emission has been observed to be overluminous with respect to HCN in Arp\,220 \citep{Huettemeister1995,Cernicharo2006,Aalto2007}.
A similar relative enhancement of HNC is only observed
in Mrk\,231, with a dominant AGN, and NGC\,4418, with a putative buried AGN, which supports the high HNC/HCN ratio to be a chemical indicator
of an X-ray dissociation region \citep[XDR,][]{Aalto2007}.
The two nuclei of Arp\,220 are revealed to show different physical conditions as observed from the high resolution maps of HNC \citep{Aalto2009}.
Model calculations attribute the observed H$_3$O$^+$ emission to enhanced X-ray irradiation \citep{vanDerTak2008}.
In the FIR, however, the lower level absorption lines in the spectrum of Arp\,220 match those observed in the diffuse clouds in the envelope of the
%front of the star-forming
Sgr~B2 molecular cloud complex hosting a massive star forming event \citep{Gonz'alez-Alfonso2004}.

Although the molecular line observations in Arp\,220 are still limited to the brighter species,
the compilation of molecular transitions by \citet{Greve2009} as well as the detection of complex organic species \citep{Salter2008} show that this galaxy
is one of the brightest molecular emitters outside the Galaxy together with the starbursts galaxies NGC\,253 and M\,82.
This fact turns Arp\,220 into a well
%Similar to those starbursts galaxies, Arp\,220 is the best 
suited candidate for molecular line surveys.

It is clear that no matter which powering source drives 
%the heating responsible for 
the large IR luminosity in Arp\,220, it will certainly have
an imprint on the physical properties and chemical composition of the ISM in this galaxy.
In this paper we present the first unbiased chemical study in the 1.3~mm spectral band of Arp\,220 with the aim of finding additional physicochemical
clues on the nature of its hidden power source.

\section{Observations and calibration}

\begin{figure*}
\centering
\includegraphics[angle=-90,width=17cm]{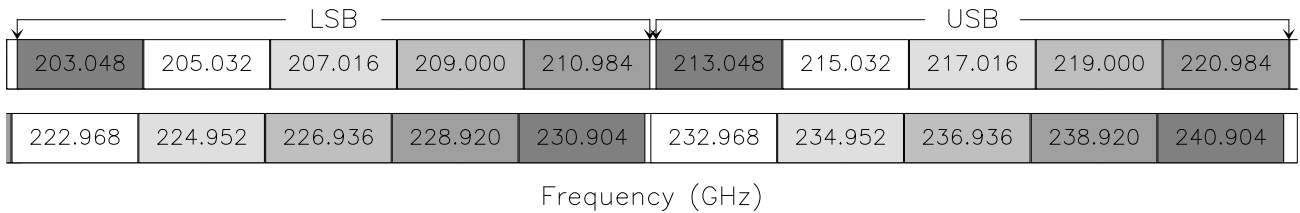}
\caption{Summary of the frequency coverage of the line survey between 202\,GHz and 242\,GHz.
The central frequency of each band is indicated.
%Similar grey scale indicate simultaneous observation in the lower (LSB) and upper (USB) sideband as indicated.
The pair of lower (LSB) and upper (USB) sidebands in the observations are indicated by shading the same grey scale.
The gap between observed bands is 16\,MHz but for the two frequencies shown in the center of the graphic
(at 212.016 and 231.936\,GHz) where the coverage gap width is 96\,MHz.\label{fig.coverage}}
\end{figure*}

Observations were carried out with the Submillimeter Array
(SMA) in Mauna Kea, Hawaii.
A total of 40\,GHz of the 1.3\,mm spectrum of Arp\,220 were covered between the frequencies of 202 and 242\,GHz in the rest frame.
This range covers the $\sim30\%$ of the 1.3\,mm atmospheric window
%with the higher atmospheric transparency, 
and the upper $71\%$ of the SMA 230\,GHz receivers nominal tunning range.

The correlator configuration provided an instantaneous 1.968\,GHz bandwidth with a 0.8125\,MHz resolution
in each sideband.
Using both sidebands, with a separation of 10\,GHz, yielded 3.936\,GHz per frequency setup.
To optimize the frequency coverage, adjacent tuning setups were spaced by 1.984\,GHz so that the gap between the observed bands
was 16\,MHz ($\sim 21$\,\kms) except for the frequencies 212.016 and 231.936\,GHz where a gap of 96\,MHz ($\sim127$\,\kms) was not covered.
With the aim of minimizing the telescope time required to complete the survey we spent only half a track per frequency setup so a total of
$\sim 8$\,GHz could be observed per night.
In Fig.~\ref{fig.coverage} we show a summary of the frequency tuning setups as well as the frequency coverage.

The phase reference center of the observations was $\alpha_{J2000}=15^{\rm h}34^{\rm m}57\fs10$ and $\delta_{J2000}=+23\degr30\arcmin11\farcs5$ with
frequencies redshifted to a velocity of $V_{\rm LSR}=5450$\kms.
Molecular emission towards Arp\,220 is extended over $\sim 2''-4''$ in size as observed in the CO $J=1-0$ \citep{Downes1998}, 
$J=2-1$ \citep{Sakamoto1999,Downes2007}, and $J=3-2$ \citep{Sakamoto2008}.
We used either compact or subcompact SMA configurations so that the vast majority of this emission would lie within the synthesized beam.

\begin{table*}
\begin{center}
\small
\caption{Observational details of each frequency setup\label{tab.obsdetails}}
\begin{tabular}{cllllll}
\hline
\hline
Date              &   Frequency            & Config. \tablefootmark{a}  & $\tau_{225}$  \tablefootmark{b} &   Flux        &    Gain            &   Bandpass           \\
                  &    (BAND)              & (\# Ant.)                  &                                 &               &                    &                      \\
\hline
2007 May 06       &   215.032\,GHz (USB)   & SC (7)                     & 0.10                            &   Titan       &    J1613+342        &   3c273              \\
                  &   217.016\,GHz (USB)   & SC (6)                     & 0.09                            &   Uranus      &    J1613+342        &   Uranus             \\
2007 May 07       &   219.000\,GHz (USB)   & SC (5)                     & 0.05                            &   Callisto    &    J1613+342        &   3c273              \\
2007 May 08       &   220.984\,GHz (USB)   & SC (5)                     & 0.05                            &   Callisto    &    J1613+342        &   3c273              \\
2008 Feb 26       &   213.048\,GHz (USB)   &  C (7)                     & 0.16                            &   Titan       &    J1504+104        &   3c273              \\
                  &   222.968\,GHz (LSB)   &  C (8)                     & 0.17                            &   Callisto    &    J1504+104        &   J1924-292, Callisto \\
2008 Mar 03       &   224.952\,GHz (LSB)   &  C (8)                     & 0.19                            &   Titan       &    J1504+104        &   3c273              \\
                  &   226.936\,GHz (LSB)   &  C (8)                     & 0.20                            &   Titan       &    J1504+104        &   J1924-292, Callisto \\
2008 Mar 04       &   222.968\,GHz (LSB)   &  C (8)                     & 0.21                            &   Titan       &    J1504+104        &   3c273, Titan       \\
2008 Mar 05       &   228.920\,GHz (LSB)   &  C (7)                     & 0.20                            &   Titan       &    J1504+104        &   3c273, Titan       \\
                  &   230.904\,GHz (LSB)   &  C (7)                     & 0.20                            &   Callisto    &    J1504+104        &   3c273, Callisto, J1924-292 \\
\hline
\end{tabular}
%% Any table notes must follow the \end{tabular} command.
\tablefoot{
\tablefoottext{a}{Array configuration (SC: Subcompact - C: Compact) and number of antennae available.}
\tablefoottext{b}{Average zenith opacity at 225\,GHz during observations.}
}
\end{center}
\end{table*}

Gain calibration consisted of 4\,min observing scans of the quasars J1613+342 or J1504+104
at a distance of $13.6^\circ$ and $14.9^\circ$ from Arp\,220, respectively, which were sampled every $\sim20$ min.
The details on the observation of each individual frequency setup are summarized in Table~\ref{tab.obsdetails} in which the observing date,
SMA configuration, number of antennae available, weather conditions, and the calibrators used are given.
Table~\ref{tab.caldetails} presents some parameters of the resulting clean maps for each observed frequency.
We achieved an average rms of $22$\,mJy/beam on a 6.5\,MHz channel width (i.e. 8 averaged original channels) with synthesized beams ranging from
the smallest ($2.9''\times2.2''$) to the largest ($10''\times6.9''$) resulting from compact and subcompact configurations, respectively.

Data reduction and calibration were performed using the \textsc{MIR-IDL} package and imaging was carried out using \textsc{MIRIAD}.

\begin{table*}
\begin{center}
\caption{Resulting image parameters for each observed frequency\label{tab.caldetails}}
\begin{tabular}{c c c @{$\times$} c @{$@\,$} c c c c}
\hline
\hline
Frequency            & $rms_{1\sigma}$\,\tablefootmark{a} &       \multicolumn{3}{c}{Synt. Beam}               &  $S_\nu$\,\tablefootmark{b} &  T$_{\rm mb}$/S\,\tablefootmark{c} & T$_{\rm b}$/S\,\tablefootmark{d} \\
(GHz)                &      (mJy/beam)                    & $\rm (B_{maj}''$ & $\rm B_{min}''$  & P.A$^\circ$) &  (mJy)                      &  K/Jy                               & K/Jy          \\
\hline
203.048              &       19.7                       &   $ 3.9 $         &  $ 3.0 $         &  $+82$        &  152                        &  2.54                               &  7.42         \\
205.032              &       23.5                       &   $ 8.8 $         &  $ 7.3 $         &  $+76$        &  135                        &  0.45                               &  7.27         \\
207.016              &       21.0                       &   $ 8.8 $         &  $ 6.2 $         &  $+76$        &  143                        &  0.52                               &  7.13         \\
209.000              &       22.0                       &   $ 7.6 $         &  $ 5.0 $         &  $+76$        &  175                        &  0.73                               &  7.00         \\
210.984              &       20.4                       &   $10.0 $         &  $ 6.9 $         &  $+47$        &  177                        &  0.40                               &  6.86         \\
213.048              &       21.2                       &   $ 3.8 $         &  $ 2.8 $         &  $+82$        &  145                        &  2.51                               &  6.74         \\
215.032              &       24.4                       &   $ 8.5 $         &  $ 7.0 $         &  $+77$        &  156                        &  0.44                               &  6.61         \\
217.016              &       21.9                       &   $ 8.4 $         &  $ 6.1 $         &  $+77$        &  178                        &  0.50                               &  6.49         \\
219.000              &       23.3                       &   $ 7.3 $         &  $ 4.8 $         &  $+76$        &  255                        &  0.73                               &  6.37         \\
220.984              &       21.5                       &   $ 9.5 $         &  $ 6.7 $         &  $+47$        &  218                        &  0.40                               &  6.26         \\
222.968              &       17.6                       &   $ 2.9 $         &  $ 2.6 $         &  $-49$        &  145                        &  3.22                               &  6.15         \\
224.952              &       23.3                       &   $ 3.6 $         &  $ 2.9 $         &  $-49$        &  165                        &  2.31                               &  6.04         \\
226.936              &       20.6                       &   $ 2.9 $         &  $ 2.6 $         &  $-54$        &  202                        &  3.07                               &  5.93         \\
228.920              &       23.8                       &   $ 3.2 $         &  $ 2.4 $         &  $-62$        &  194                        &  3.02                               &  5.83         \\
230.904              &       23.9                       &   $ 2.9 $         &  $ 2.3 $         &  $-61$        &  503                        &  3.44                               &  5.73         \\
232.968              &       18.1                       &   $ 2.9 $         &  $ 2.4 $         &  $-47$        &  153                        &  3.17                               &  5.63         \\
234.952              &       15.7                       &   $ 3.7 $         &  $ 2.8 $         &  $-41$        &  177                        &  2.16                               &  5.54         \\
236.936              &       22.7                       &   $ 2.9 $         &  $ 2.4 $         &  $-50$        &  204                        &  3.03                               &  5.44         \\
238.920              &       25.8                       &   $ 3.1 $         &  $ 2.4 $         &  $-54$        &  205                        &  2.91                               &  5.35         \\
240.904              &       25.9                       &   $ 2.9 $         &  $ 2.2 $         &  $-55$        &  216                        &  3.30                               &  5.26         \\
\hline
\end{tabular}
%% Any table notes must follow the \end{tabular} command.
\tablefoot{
\tablefoottext{a}{Calculated for 6.5\,MHz channels.}
\tablefoottext{b}{Average flux density measured in the whole $\sim2$\,GHz band.}
\tablefoottext{c}{Flux density to synthesized main beam brightness temperature under the point source assumption.}
\tablefoottext{d}{Flux density to source brightness temperature assuming a $2''$ Gaussian emitting region.}
%\tablefoottext{b}{Average zenit opacity at 225\,GHz during observations.}
}
\end{center}
\end{table*}

\subsection{Flux calibration and bandpass determination}
\label{sect.calibration}
Absolute flux density calibration was derived individually for each track from observations of
Titan, Uranus or Callisto as indicated in Table~\ref{tab.obsdetails}.
The nominal absolute calibration accuracy achieved with the SMA is $\sim15\%$.
As explained above, observations were carried out so that there were no overlaps among adjacent bands, so
the relative flux density calibration becomes more important.
To account for variations of the individual flux density calibrations on each of the observed frequency bands
we used the flux densities measured for the gain calibrators and fitted their emission dependance with frequency
assuming a power law dependance as $S_\nu\propto \nu^\alpha$.
Given that the observations were all carried out within a short period of time (see Table~\ref{tab.obsdetails})
we can assume the flux density of the quasars not to vary significantly ($\la10\%$).
We fitted power laws with $\alpha=-1.1$ and $-1.6$ for J$1504+104$ and J$1613+342$, respectively.
These power laws are similar to those derived from the upper and lower sidebands on individual observations,
which supports the assumption of little daily variation on their flux densities.
Corrections based on deviations of the quasar flux densities from the fitted power law were applied to the Arp\,220 data.
These corrections were smaller than $13\%$ in absolute flux density in all cases.
From these corrections we estimate an accuracy in the relative flux density determination of $\sim6\%$ across the 40\,GHz observed band.

Bandpass calibration was derived from long integrations on 3c273, Uranus, or J$1924-292$.
However, the integrations on Callisto and Titan were included in some cases (see Table~\ref{tab.obsdetails}).
Titan shows a strong CO line which appears at $\sim235$\,GHz and thus it was not used for bandpass in this setup.
The accuracy of the bandpass determination was checked on the observed quasars, J$1613+342$ or J$1504+104$, to look for possible bandpass residuals
which might mimic line emission. No significant bandpass glitches were found that might affect the analysis presented in this paper.
We used this CO line contamination on Titan to check the consistency of the relative flux density calibration based on the assumption
of constant flux density of the quasars.
We wrongly calibrated the data without masking the channels where CO contributes to the total flux measured in Titan
and subsequently applied the correction derived from the power law fitted to J$1504+104$ density fluxes.
The resulting corrected flux matched within $<5\%$ the calibrated data derived from Titan with the masked CO line.

%The resulting difference in the calibration is consistent with that derived based on the assumption of constant flux density of J$1504+104$,
%as detailed in the previous paragraph.
%By applying the relative flux calibration, based on the flux variations of $J1504+104$, to the data with CO contamination in Titan,
%we were able to accurately recover the absolute flux derived from the data calibrated with the masked Titan CO emission.

\section{Continuum emission}
\label{sect.Continuum}

\begin{table}
\begin{center}
\caption{Average continuum fluxes in selected line-free spectral bands\label{tab.ContFluxes}}
\begin{tabular}{c c c c c c}
\hline
\hline
Frequency            &    Flux       & $rms$    &   $\#$\,Channels \tablefootmark{a} \\
Range                &   density     &          &                                    \\  
(GHz)                &    (mJy)      & (mJy)    &                                    \\
\hline
$203.585-203.739$    &    130.3      & 28    &  31    \\
$203.775-204.528$    &    130.6      & 35    &  148   \\
$206.420-206.963$    &    132.6      & 38    &  107   \\
$208.501-208.731$    &    134.1      & 58    &  46    \\
$212.552-212.701$    &    123.3      & 33    &  30    \\
$213.643-213.894$    &    132.3      & 46    &  50    \\
$213.936-214.151$    &    142.0      & 31    &  43    \\
$214.561-214.751$    &    148.4      & 32    &  38    \\
$222.465-223.146$    &    136.6      & 31    &  134   \\
$223.582-223.751$    &    129.0      & 29    &  34    \\
$224.674-225.196$    &    147.9      & 27    &  103   \\
$231.409-231.849$    &    153.6      & 41    &  87    \\
$232.128-233.470$    &    149.6      & 29    &  263   \\
$234.599-235.490$    &    171.9      & 36    &  175   \\
$240.063-240.662$    &    178.6      & 46    &  118   \\
\hline
\end{tabular}
%% Any table notes must follow the \end{tabular} command.
\tablefoot{
\tablefoottext{a}{Number of averaged 5.1\,MHz channels to estimate the continuum flux.}
%\tablefoottext{b}{Average zenit opacity at 225\,GHz during observations.}
}
\end{center}
\end{table}

\begin{figure*}
\centering
\includegraphics[angle=-90,width=17cm]{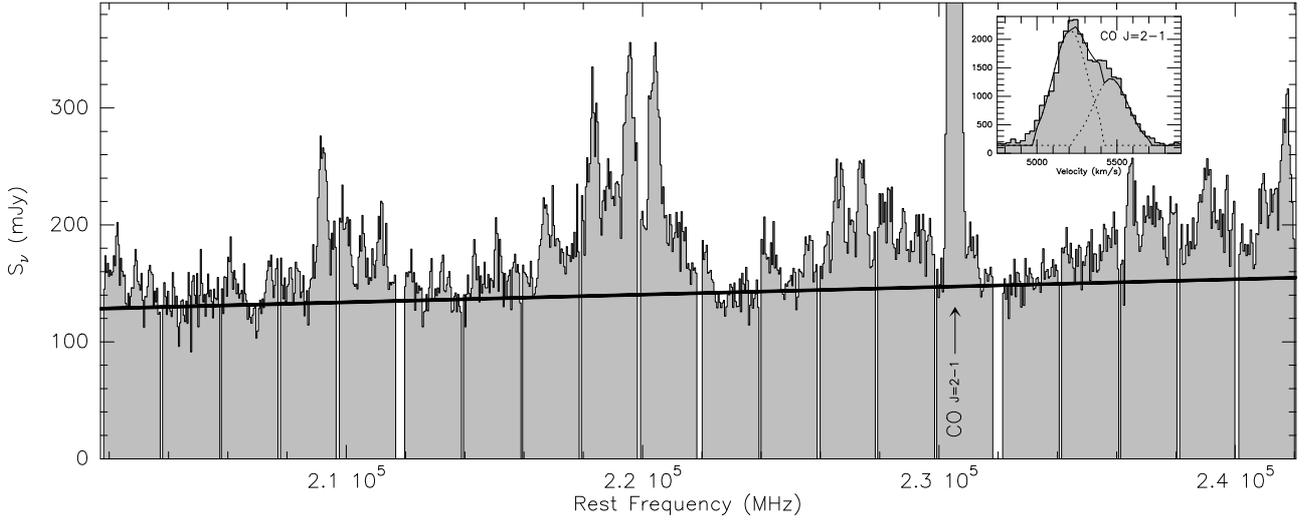}
\caption{Composite of the 1.3\,mm frequency scan carried out with the SMA between 202 and 242\,GHz.
Each individual spectrum was extracted from the natural weighted datacubes at the peak of emission position
$\alpha_{J2000}=15^{\rm h}34^{\rm m}57\fs22$ and $\delta_{J2000}=+23\degr30\arcmin11\farcs5$.
The calculated continuum level is represented by a straight line (see Sect.~\ref{sect.Continuum} for details).
The CO $J=2-1$ emission line, labeled in the Figure, dominates the spectral range.
This line is shown in a separate box in the upper right hand size, where the velocity scale in the X-axis is referred to the rest frequency of the transition.
The double Gaussian profile fitted to the CO feature (see Sect.~\ref{sect.CO} for details) are shown. both the individual velocity component (dotted line)
and the total fitted profile (solid line) are shown.
\label{fig.specJy}}
\end{figure*}

Due to broad spectral features and the prolific molecular line emission in Arp\,220, most of the observing frequency setups showed little or
no line free frequency ranges to be used to subtract the continuum from individual setups.
Thus, continuum emission was not subtracted in the UV data.
Fig.~\ref{fig.specJy} shows a composite spectrum of the whole covered 1.3\,mm band.
This figure has been composed with the individual spectra extracted from the peak position of the unresolved emission at
$\alpha_{J2000}=15^{\rm h}34^{\rm m}57\fs22$ and $\delta_{J2000}=+23\degr30\arcmin11\farcs5$.

\begin{figure}
\resizebox{\hsize}{!}{\includegraphics[angle=-90]{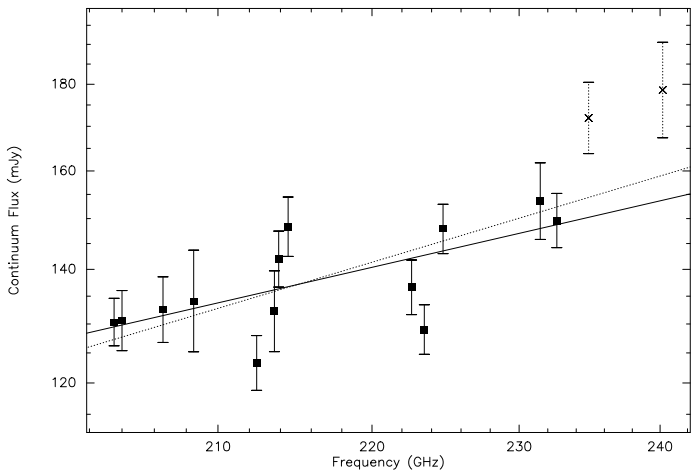}}
\caption{Power law fit to the continuum measurements in molecular emission free bands across the whole 40\,GHz surveyed (see Table~\ref{tab.ContFluxes}).
Fit to all measurements (dotted line) result in a dependance as $S_\nu\propto\nu^{1.3}$.
The final fit applied to the data (solid line), as derived from the measurements represented by filled squares, shows a power law dependance
as $S_\nu\propto\nu^{1.0}$.
Error bars represent the rms with respect to the average within each selected frequency range.
See text in Sect.~\ref{sect.Continuum} for details.
\label{fig.Baseline}}
\end{figure}

Table~\ref{tab.caldetails} shows the average flux densities measured in each individual observed $\sim 2$\,GHz band, where line emission was not subtracted.
In order to estimate and subtract the continuum emission from the spectra we selected a total of 15 narrow frequency bands that we considered as free of molecular emission.
The positions of these bands and the measured continuum flux densities are presented in Table~\ref{tab.ContFluxes}. There we also present the measured rms with respect to the
mean flux density and the number of 5.1\,MHz channels ($\sim 6.4-7.5$\,\kms) averaged for each band.
These values are plotted in Fig.~\ref{fig.Baseline} where their associated error bars are derived from the measured rms around the mean flux density value.
The total frequency range covered appears to be large enough to show the changes in the continuum flux as clearly depicted in the Fig.~\ref{fig.Baseline}.
We fitted a power law dependance to the measurements as $S_\nu\propto\nu^{\alpha}$
with each measurement weighted as $(rms/\sqrt{n})$ where $n$ is the number of averaged channels.
A power law index of $\alpha=1.3\pm0.2$ (dotted line in Fig.~\ref{fig.Baseline}) is derived by fitting all the selected bands in Table~\ref{tab.ContFluxes}.
However, we noticed that the flux density measurements at the two highest frequency continuum bands significantly disagree ($>10\%$) with the fitted frequency dependance.
Such a discrepancy might indicate that the measured flux densities at the two highest frequencies may be overestimated due 
to a significant line contamination or a significant error in the relative flux density calibration.
%line contamination at these frequencies or an error in the relative flux calibration.
The solid line in Fig.~\ref{fig.Baseline} represents the fit to the data excluding the values at the highest frequencies, which results in a slope of
$\alpha=1.04\pm0.02$.
The difference in flux density between both fits is $<5\%$ for the frequency range covered in our observations, which is within our calibration accuracy (Sect.~\ref{sect.calibration}).
The continuum flux densities measured around 212.6, 214.6, and 223.6\,GHz also show deviations from both fits, but the differences from either of the fits
are still within $<10\%$.
Finally, we adopted the power law $\nu^{1.0}$ as the frequency dependence for the continuum flux densities to subtract it from the spectra.

The average continuum flux in the 1.3\,mm band is $142$\,mJy, ranging from 129\,mJy to 154\,mJy at 203 and 241\,GHz, respectively.
%Fig.~\ref{fig.ContCO} shows the continuum emission as imaged from the molecular emission free channels in the 230.904\,GHz setup.

\subsection{Line contribution to total emission flux in broad band observations}
\label{sect.line2cont}
The data set presented in this work provides an unique opportunity to accurately estimate the contamination of line emission to the continuum flux density
measured by broad band detectors at millimeter wavelengths.
The average total flux recovered in the 40\,GHz spectral band covered by the survey is $195$\,mJy.
As seen in Fig.~\ref{fig.specJy} small subpanel, the molecular line emission at these frequencies is dominated by the $J=2-1$ transition of carbon
monoxide. However, the contribution from other molecular species (see Sect.~\ref{sect.Molecular}) is far from being negligible.
Assuming the power law spectrum fitted in the previous section as the continuum level, we 
assess the total line flux of both CO and other species by integrating the continuum subtracted spectra.
%calculated the observed flux of both CO and other species as the integrated flux of the continuum subtracted spectra.
We find that the contribution of molecular emission to the overall 40\,GHz measured flux constitutes $28\%$, from which only a $9\%$ of the total flux is
attributed to the CO $J=2-1$ line.
The work by \citet{Albrecht2007} estimated an average contamination by CO to the 1.3\,mm flux density of $21.3\%$ based
on 1.3\,mm bolometer ($\sim$50~GHz bandwidth) and CO $J=2-1$ observations in a sample of 99 galaxies.
Although their sample does not include ULIRGs, where line contribution appears to be very important, the emission of molecules other than CO might be also
important in some of their sources. In particular, the line contamination to the continuum flux density might become significant in starbursts,
where an also prolific molecular emission is observed \citep{Mart'in2006}.

Our observations show that at short mm and sub-mm wavelengths, the flux density correction due to molecular line emission is more important than 
the non-thermal contribution extrapolated from radio measurements \citep{Norris1988,Scoville1997}.
For Arp\,220 the non-thermal contribution is estimated to be 7.7\,mJy (203\,GHz) and 6.7\,mJy (241\,GHz), equivalent to $6\%$ and $3\%$ of the averaged total flux
at those frequencies, respectively, and well below the $28\%$ line contribution.

%CS 5-4 in BOLOMETERS???

\subsection{Comparison with previous measurements}
\citet{Scoville1997} and \citet{Sakamoto1999} reported continuum flux densities of $192\pm20$\,mJy and $208$\,mJy, respectively,
in a 1\,GHz band centered at 229.4\,GHz with the OVRO interferometer.
Within the same frequency range we derived an averaged flux density of $185\pm3$\,mJy, in good agreement with their measurement.
%We only considered the statistical error in this measurement.
However, at these frequencies, after subtracting the line contribution of $\sim 39$\,mJy ($\sim21\%$ of the total flux density),
the effective continuum flux yields $146$\,mJy.
The continuum flux density would be therefore overstimated by $\sim30\%$ due to the contamination from line emission.

\citet{Woody1989} measured a flux density of $140\pm20$\,mJy at a sky frequency of 219.5\,GHz also with the OVRO interferometer.
Luckily this 1\,GHz band falls into the 
%redshifted 
$223.05-224.07$\,GHz rest frequency band where our survey shows very little line contamination.
Our flux density in that range is $146\pm3$\,mJy, with just an estimated $2\%$ of line contribution.
In this case, the corrected continuum flux density we measure of $142$\,mJy agrees with that by \citet{Woody1989}.

The agreement between the flux density measured in this work and that of \citet{Scoville1997},
with a beam $\sim6.5$ times smaller in area, implies that most of the 
flux density is recovered in both the SMA and OVRO interferometric observations.
The discrepancy between the observed flux density by \citet{Woody1989} and the single dish flux density of $340\pm80$\,mJy by \citet{Thronson1987} is likely due to
strong line contamination in the frequency tuning of the latter (not specified in the publication). Similarly, our observations
can easily explain the average single dish flux of $226\pm10$\,mJy measured in a broad window centered at 240\,GHz \citep{Carico1992}, where
molecular line emission such as that of CS $J=5-4$ and even the $J=3-2$ transitions of HCO$^+$ and HCN, among many others,
will strongly contribute to the continuum emission.

\section{Molecular emission}
\label{sect.Molecular}

\subsection{Carbon Monoxide}
\label{sect.CO}

\begin{figure}
\resizebox{\hsize}{!}{\includegraphics[angle=-90]{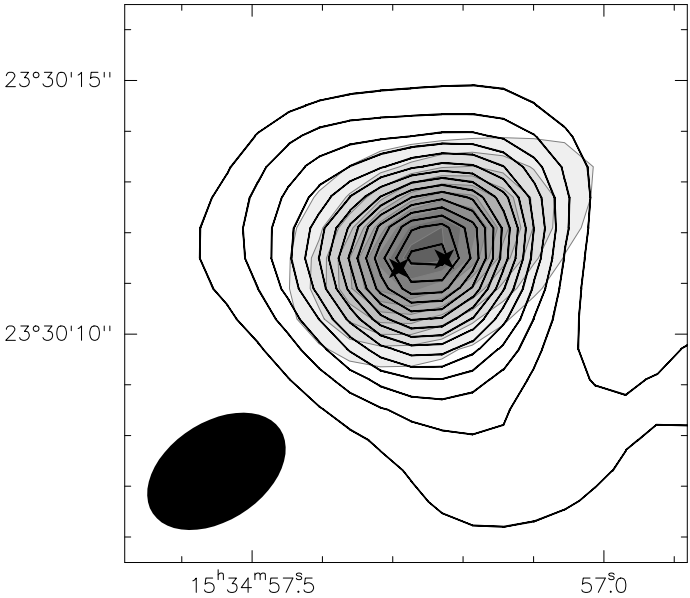}}
\caption{Robust weighted images of the continuum emission (grey scale) and CO $J=2-1$ integrated flux (contours) at a resolution of $2.9''\times1.9''$.
Continuum emission has been estimated from the 230.904\,GHz setup in the ranges not affected by CO or $^{13}$CS.
The stars represent the position of the 1.3\,mm continuum sources from the high resolution images of \citet{Sakamoto1999}.
Contours are $3\sigma$ significant levels of 14\,mJy\,beam$^{-1}$ and 49\,Jy\,km\,s$^{-1}$ for the continuum and line flux, respectively.
% TOTAL FLUX Cont 183 mJy
\label{fig.ContCO}}
\end{figure}

\begin{table}
\begin{center}
\caption{CO $J=2-1$ fitting parameters \label{tab.COfit}}
\begin{tabular}{c @{\quad} c @{\quad} c @{\quad} c @{\quad} c @{\quad} c}
\hline
\hline
         &  $\int{S_\nu\,{\rm d}v}$   &    $\int{T_{\rm B}{\rm d}v}$ & $V_{\rm LSR}$             &  $\Delta v_{1/2}$ \tablefootmark{a}  & $T_{\rm B}$ \\
         &  Jy\,km\,s$^{-1}$         &     K\,km\,s$^{-1}$          & km\,s$^{-1}$              &  km\,s$^{-1}$                        & K           \\
\hline
1            &  $548\pm5$                &    $3140\pm30$               & $5208.4\pm1.2$            &  $254.6\pm1.6$                   & 11.6   \\ %   & 11.583       \\
2            &  $317\pm3$                &    $1820\pm20$               & $5461.7\pm2.0$            &  $254.6\pm1.6$                   & 6.7    \\ %   & 6.708       \\
\hline
\end{tabular}
\tablefoot{These parameters are used to constrain the LTE analysis of all other molecular species. See text in Sect.~\ref{sect.LTE} for details.
\tablefoottext{a}{Double Gaussian fit with equal linewidth.}
}
\end{center}
\end{table}

\begin{figure*}
\centering
\includegraphics[width=17cm]{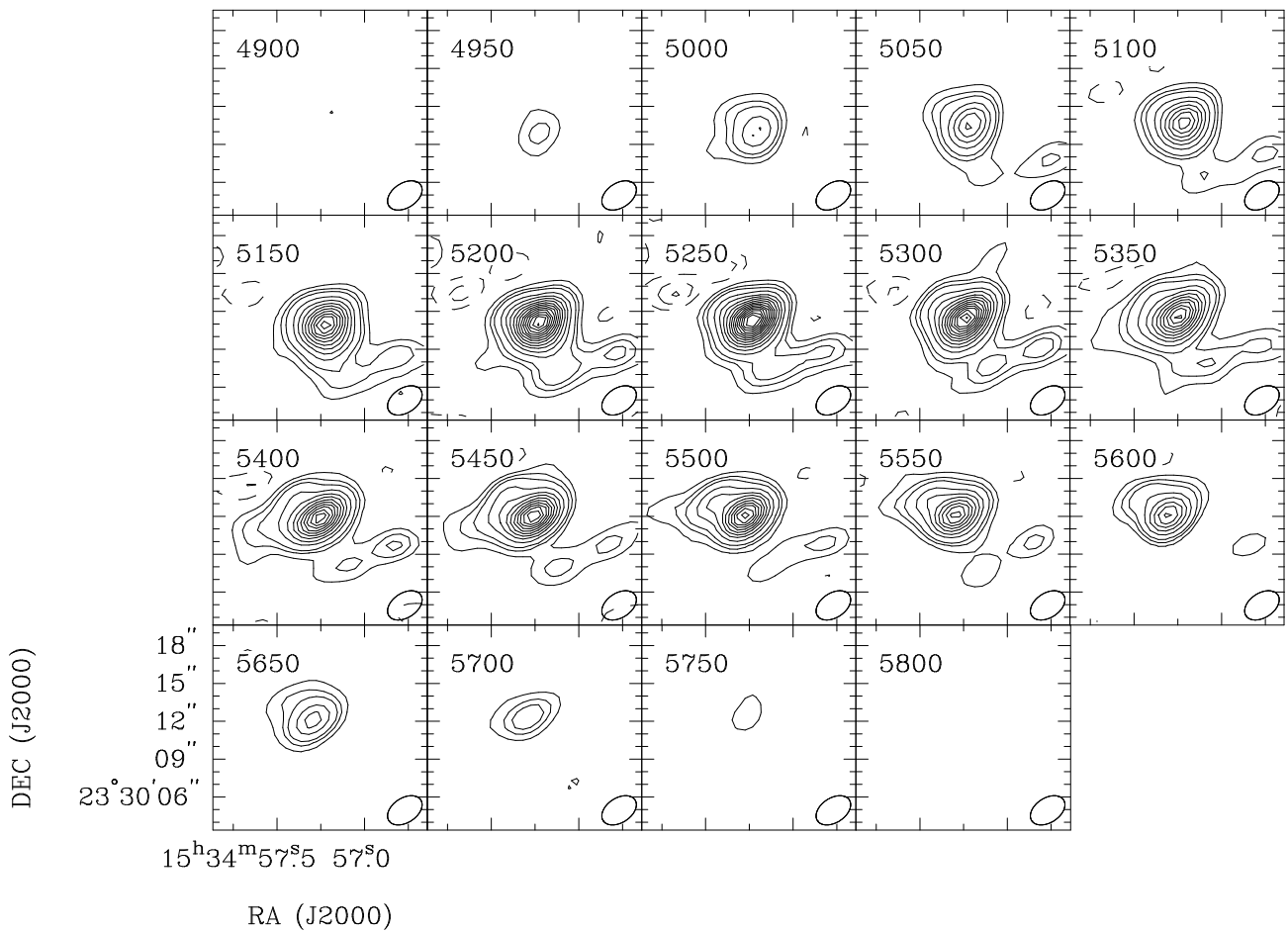}
\caption{CO $J=2-1$ channel maps in steps of 50\,km\,s$^{-1}$.
First 3 contours are $3\sigma$ significant and then $6\sigma$ up to $100\sigma$, with $\sigma=19$\,mJy.
The number at top-left in each panel denotes the LSR velocity in \kms.
\label{fig.COchan}}
\end{figure*}

The subpanel in Fig.~\ref{fig.specJy} shows the prominent emission of $^{12}$CO $J=2-1$ with a peak flux of 2.3\,Jy.
The overall linewidth derived from fitting a single Gaussian to the line profile is $419.3\pm1.9$\,km\,s$^{-1}$, in agreement with that measured by \citet{Greve2009}
with the JCMT telescope.
Within the synthesized beam we measure an integrated flux of $\sim 865$\,Jy\,km\,s$^{-1}$.
Fortunately, near the CO line there are apparently line-free channels within the observed band, and therefore, continuum subtraction
could be easily done in the UV-plane.
Fig.~\ref{fig.ContCO} shows the continuum subtracted integrated CO emission overlaid on top of the line-free continuum emission.
The total recovered CO flux in the map is $1650\pm50$\,Jy\,km\,s$^{-1}$, in good agreement with previous single-dish JCMT
measurements \citep{Wiedner2002,Greve2009}.
Thus we recover all the single-dish flux in our maps.
This integrated flux is $30\%$ larger than that recovered by the higher angular resolution maps of OVRO \citep[$1.2''\times1.0''$][]{Scoville1997}
and PdBI \citep[$0.7''\times0.5''$,][]{Downes1998}, due to the missing short-spacing flux in their maps.
Fig.~\ref{fig.COchan} shows the 50\,km\,s$^{-1}$ channel maps where CO emission is detected over a 800\,km\,s$^{-1}$ range, between 4950 and 5750\,km\,s$^{-1}$.
%We detect CO emission in a 800\,km\,s$^{-1}$ range in $V_{\rm LSR}$ between 4950 and 5750\,km\,s$^{-1}$.

The spatial resolution of our observations appears not to be enough to disentangle the chemistry of the two nuclei.
However, the velocity profile provides us with some rough information about the origin of the molecular emission.
Although the broad emission from both nuclei is significantly blended in velocities,
as seen in the high resolution maps of CO $1-0$ and $2-1$ \citep{Scoville1997,Sakamoto1999}, the lower velocity component is mostly the emission
from the southern region and the western nucleus, whereas
%On the other hand, 
the higher velocity component mostly traces the emission from the northern region and the eastern nucleus.

We fitted a double Gaussian profile to the spectrum extracted from the emission peak position shown in the inset in Fig.~\ref{fig.specJy}.
The fitting assumed both velocity components to have equal linewidths.
This constraint of double Gaussian profile with an equal linewidth had little impact ($<13\%$) in the relative integrated intensities of each
component as compared to an unconstrained fitting.
The CO parameters derived from our double Gaussian fit are shown in Table~\ref{tab.COfit} and the fitted profiles are overlaid on top of the CO
spectrum in Fig.~\ref{fig.specJy}.

\subsection{Line Identification}
We used the rest frequencies provided by the JPL \citep{Pickett1998} and CDMS \citep{Muller2001,Muller2005} molecular spectroscopy catalogues to identify each spectral feature in
the survey.
As shown in Fig.~\ref{fig.specKmodel}, a total of 73 individual or groups of molecular transitions were identified
% contributing more than {\bf 1mK???} as individual or being part of the observed
%features 
in the 40\,GHz frequency band covered.
Although an average of 1.8 lines/GHz are detected, given the broad emission towards Arp\,220 of $\rm FWHM\sim330$\,MHz, as measured from CO $J=2-1$,
a significant fraction of the observed 1.3\,mm band is confusion limited.
%430\,\kms (330\,MHz)

A total of 15 molecular species and 6 isotopical substitutions were identified and fitted.
Among them CH$_2$CO is tentatively identified for the first time in the extragalactic ISM,
as well as the isotopologues $^{29}$SiO, H$_2^{18}$O, and all the $^{13}$C substitutions of HC$_3$N.
Table~\ref{tab:census} provides an updated list of all the species identified in the extragalactic interstellar medium.
A total of 46 species and 23 isotopical substitution have been identified to date.

\begin{table*}
\begin{center}
\caption{Census of extragalactic molecular species and isotopologues detected}
\label{tab:census}
%\begin{tabular}{l @{} l @{} l @{\,\,\,} l @{\,\,\,} l @{\,\,\,} l}
\begin{tabular}{ llllll }
\hline
{\bf 2 atoms}           &       {\bf 3 atoms}   &       {\bf 4 atoms}   &       {\bf 5 atoms}   &       {\bf 6 atoms}           &       {\bf 7 atoms}   \\
\hline
\hline
OH                      &       H$_2$O, {\tiny H$_2^{18}$O}
                                                &       H$_2$CO         &       c-C$_3$H$_2$    &       CH$_3$OH, {\tiny $\rm^{13}CH_3OH$} &       CH$_3$C$_2$H    \\
CO {\tiny \hspace{-5pt} $\Bigg\{ \hspace{-5pt} \begin{array}{l} ^{13}CO \\ C^{18}O \\ C^{17}O \\ \end{array} $}
                        &       HCN {\tiny \hspace{-5pt} $\Bigg\{ \hspace{-5pt} \begin{array}{l} H^{13}CN \\ HC^{15}N \\ DCN \\ \end{array} $}
                                                &       NH$_3$          &       HC$_3$N {\tiny \hspace{-5pt} $\Bigg\{ \hspace{-5pt} \begin{array}{l} H^{13}CCCN \\ HC^{13}CCN \\ HCC^{13}CN \\ \end{array} $}
                                                                                                &       CH$_3$CN                &                       \\
H$_2$, {\tiny $HD$}     &       HCO$^+$ {\tiny \hspace{-5pt} $\Bigg\{ \hspace{-5pt} \begin{array}{l} H^{13}CO^+ \\ HC^{18}O^+ \\ DCO^+ \\ \end{array} $}
                                                &       HNCO            &       CH$_2$NH        &                               &                       \\
CH                      &       C$_2$H          &       H$_2$CS         &       NH$_2$CN        &                               &                       \\
CS {\tiny \hspace{-5pt} $\Bigg\{ \hspace{-5pt} \begin{array}{l} ^{13}CS \\ C^{34}S \\ C^{33}S \\ \end{array} $}
                        &       HNC {\tiny \hspace{-5pt} $\Big\{ \hspace{-5pt} \begin{array}{l} HN^{13}C \\ DNC \\ \end{array} $}
                                                &       HOCO$^+$        &       CH$_2$CO        &                               &                       \\
CH$^+$                  &       N$_2$H$^+$, {\tiny $N_2D^+$}
                                                &       C$_3$H          &                       &                               &                       \\
CN                      &       OCS             &       H$_3$O$^+$      &                       &                               &                       \\
SO, {\tiny $^{34}SO$}   &       HCO             &                       &                       &                               &                       \\
SiO, {\tiny $^{29}SiO$} &       H$_2$S          &                       &                       &                               &                       \\
CO$^+$                  &       SO$_2$          &                       &                       &                               &                       \\
NO                      &       HOC$^+$         &                       &                       &                               &                       \\
NS                      &       C$_2$S          &                       &                       &                               &                       \\
LiH                     &       H$_3^+$         &                       &                       &                               &                       \\
CH                      &       H$_2$O$^+$      &                       &                       &                               &                       \\ 
NH                      &                       &                       &                       &                               &                       \\  
OH$^+$                  &                       &                       &                       &                               &                       \\ 
HF                      &                       &                       &                       &                               &                       \\  
\hline
\end{tabular}
\tablefoot{Updated Table from \citet{Mart'in2009a}.
For references on the first detection of each species see \citep{Mart'in2006} and the reported detections of:
HD and LiH in absorption at high redshift \citep{Combes1998,Varshalovich2001};
$\rm H_3^+$ in absorption towards the ULIRG IRAS 08572+3915 \citep{Geballe2006};
$\rm H_3O^+$ in emission towards the starburst M\,82 and the ULIRG Arp\,220 \citep{vanDerTak2008};
CH and NH in absorption towards Arp\,220 \citep{Gonz'alez-Alfonso2004};
the tentative DNC, N$_2$D$^+$, $\rm C^{33}S$, and $\rm^{13}CH_3OH$ towards NGC\,253 and the Large Magellanic Cloud \citep{Mart'in2006,Wang2009,Mart'in2009b};
$\rm H_2O^+$ in emission towards Mrk\,231 and in absorption towards M\,82 \citep{vanDerWerf2010,Weiss2010};
OH$^+$ and HF in emission towards Mrk\,231 \citep{vanDerWerf2010};
and the $^{13}$C isotopologues of HC$_3$N, H$_2^{18}$O, $^{29}$SiO, and CH$_2$CO being reported in this work.
}
\end{center}
\end{table*}

\subsection{LTE modelling and fitting of the identified lines}
\label{sect.LTE}
Similar to what is found in Galactic hot cores, the line confusion and blending 
%prevents from directly 
aggravates direct fitting of Gaussian profiles to individual spectral features.
Thus it becomes absolutely necessary to fit synthetic spectra of each identified molecular species to the observed spectrum.
Synthetic spectra of each molecule were calculated using the spectroscopic parameters in the JPL and CDMS catalogues \citep{Pickett1998,Muller2001,Muller2005}.
Assuming local thermodynamic equilibrium (LTE) excitation and optically thin emission we can calculate the integrated intensity of a given transition for
any value of column density ($N$) and excitation temperature ($T_{\rm ex}$) as described in Appendix B in \citet{Mart'in2006}.
%as
%\begin{equation}
%\int{T_{\rm}\, dv}= \frac{h c^3 A_{ul} Z}{8 \pi k \nu^2 N g_u} e^{-E_u/k T_{\rm ex}} \left(1-\frac{J_{\nu}(T_{\rm BG})}{J_{\nu}(T_{\rm ex})}\right)
%\end{equation}
Due to both blending and the relatively low signal-to-noise ratio for some species we fitted a double velocity component model with
the radial velocity and linewidth fixed to those derived from $^{12}$CO in Sect.~\ref{sect.CO}.
Therefore, only an automatic exploration in the column density and excitation temperature parameters of each velocity component was carried out to fit
the model to the observed spectra.
The fitting was performed by minimization of the rms of the residuals after subtracting the synthetic spectra from the observations.
If the uncertainty due to the constraint of double velocity components with an equal width, discussed for the $^{12}$CO line fitting, applies to other molecular lines,
%,was discussed for the $^{12}$CO line fitting.
%We estimated it might introduce 
an error in the line integrated area of individual components of $<13\%$ may be introduced in the synthetic spectral fitting, which is
lower than the uncertainty in the absolute flux density and of the same order as the statistical error derived from the Gaussian fit.
However, even if the integrated intensity of each individual component can be significantly affected,
such fitting constraints are expected to introduce only a small uncertainty of $\sim1\%$ in the overall integrated area of 
the whole spectral profile as estimated from the fit to the $^{12}$CO line.

\begin{figure*}
\centering
\includegraphics[angle=-90,width=17cm]{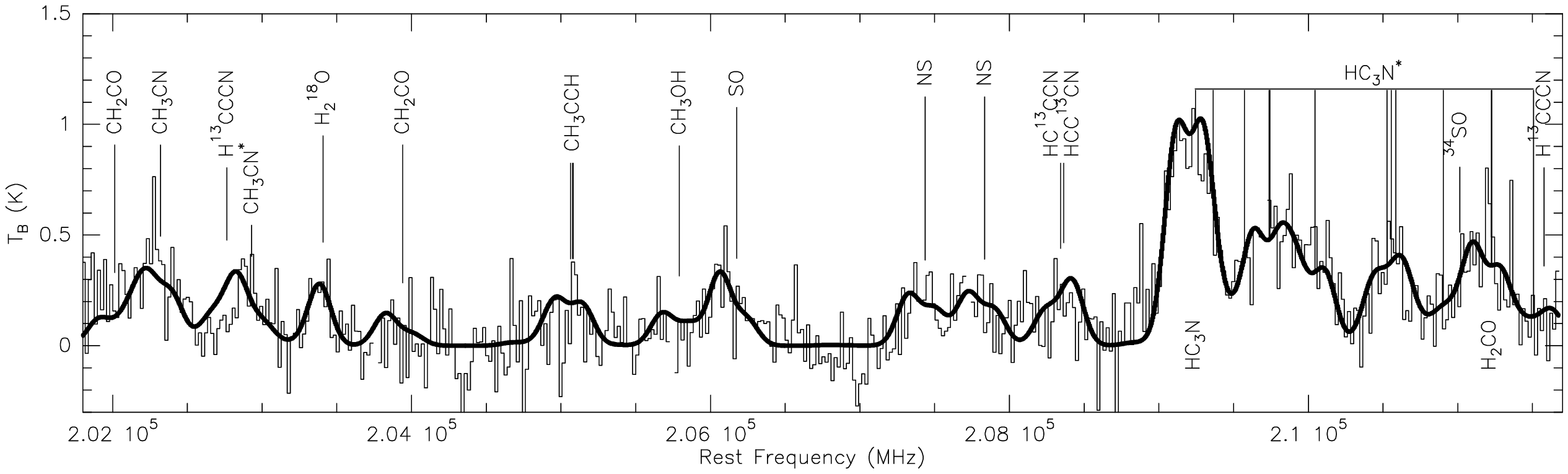}
\includegraphics[angle=-90,width=17cm]{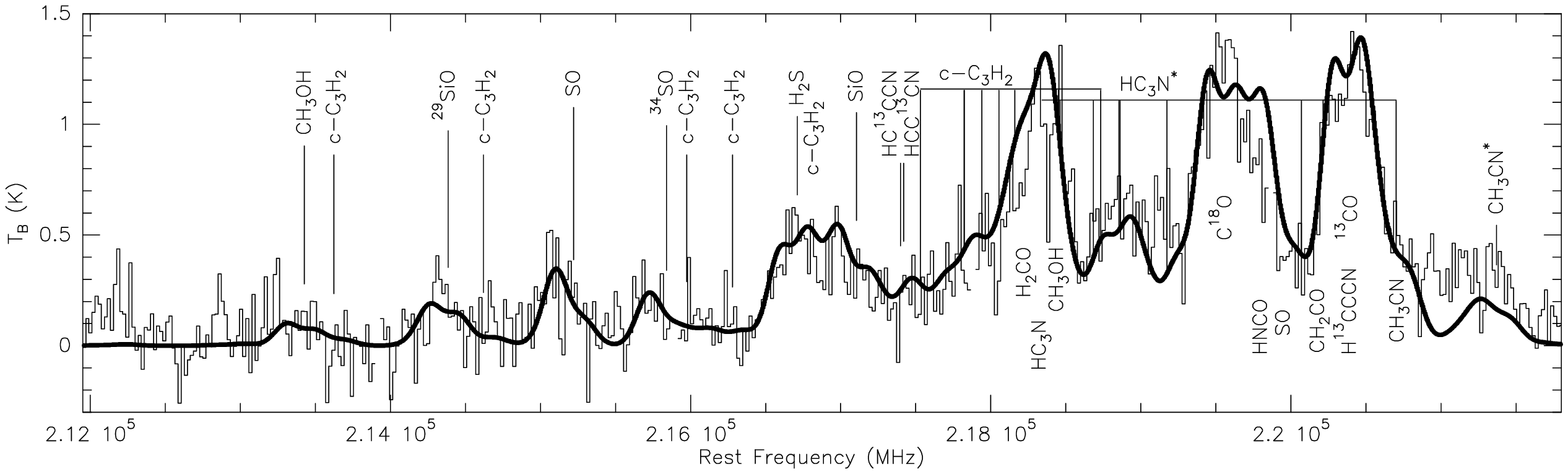}
\includegraphics[angle=-90,width=17cm]{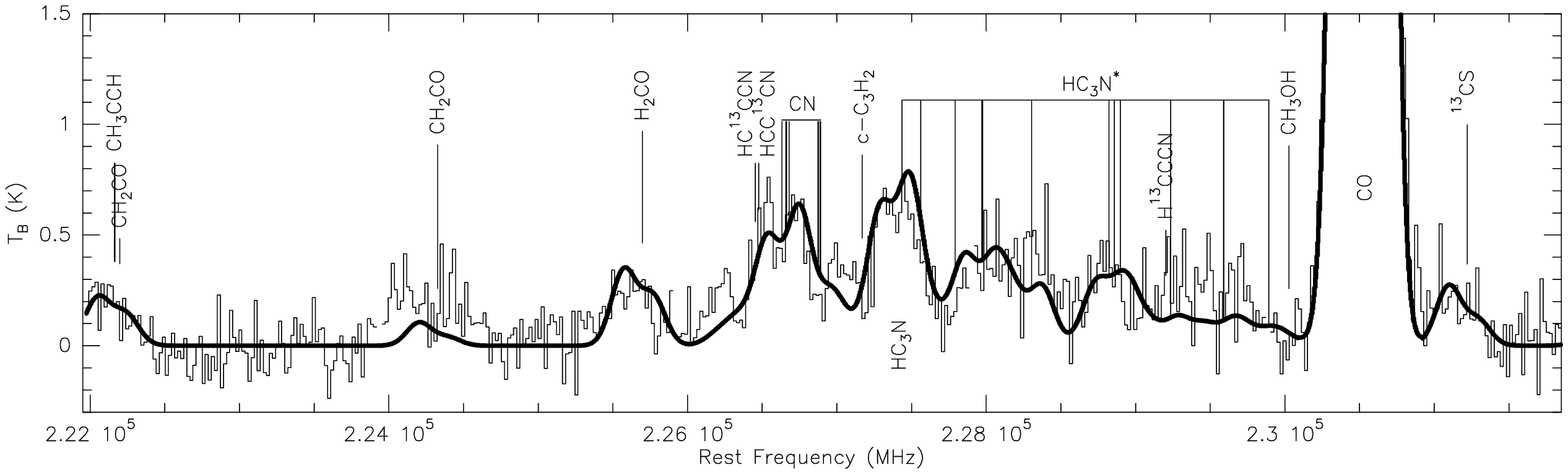}
\includegraphics[angle=-90,width=17cm]{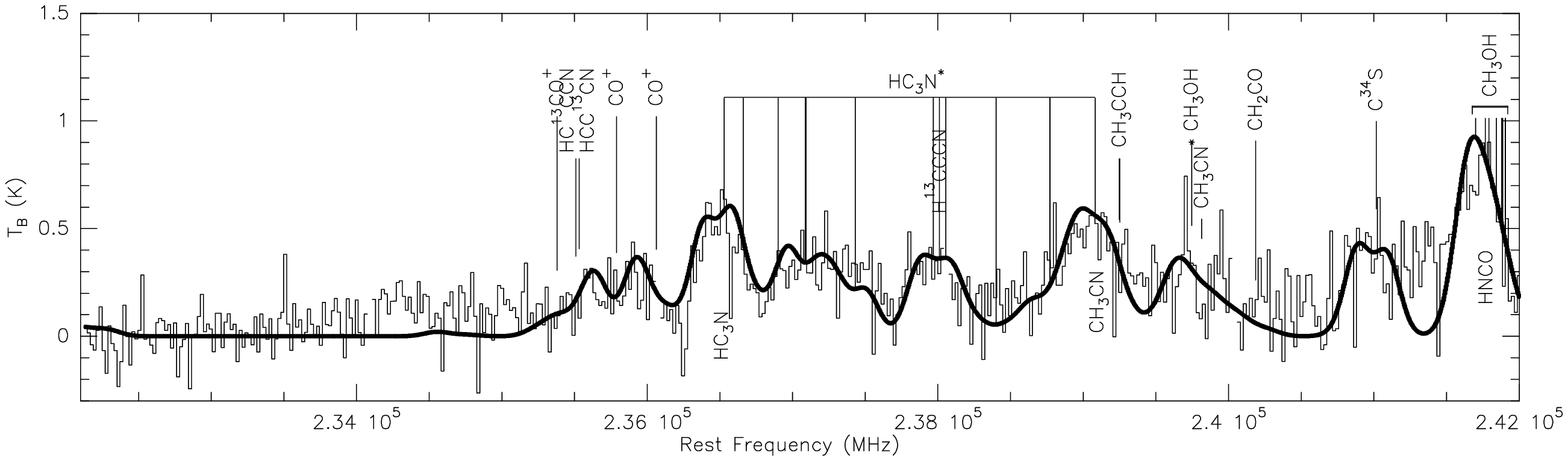}
\caption{Detailed view of the spectral line survey.
Spectral resolution of the observed data is smoothed to 20.5\,MHz ($25-30$\,\kms\,across the covered frequency range).
The LTE model of the identified molecular species is represented as a thick solid curve.
The identified molecular features are labeled.
\label{fig.specKmodel}}
\end{figure*}

As an educated cut-off to the number of lines used to generate the synthetic spectra,
we used all the listed transitions with lower level energies $E_l<1000$\,cm$^{-1}$ ($\sim1440$\,K) in the aforementioned catalogs.
Out of the $\sim 3000$ transitions used in the model fitting as shown in Fig.~\ref{fig.specKmodel}, only a limited number contribute significantly
to the observed spectrum.
Table~\ref{tab.ModelFit} shows the fitted parameters for the subset of 163 molecular transitions with a peak temperature $>25$\,mK, which corresponds
to the average $1\sigma$ rms noise level at a resolution of $\sim250$\,\kms.
These parameters are given for both velocity components centered at 5208 and 5462\,\kms, which we label as component
1 and 2, respectively.

Both Fig.~\ref{fig.specKmodel} and Table~\ref{tab.ModelFit} are scaled in brightness temperature ($T_{\rm b}$).
As detailed by \citet{Downes1989}, the conversion factor from flux density ($S$) into temperature scale can be calculated as
\begin{equation}
\label{eq.mbtojy}
%S/T_{\rm mb} =\frac{2 k}{\lambda^2} \Omega_{s \star mb}
T_{\rm mb}/S =\frac{\lambda^2}{2 k} \Omega_{s \star mb}^{-1}
\end{equation}
where $T_{\rm mb}$ is the synthesized main beam brightness temperature and $\Omega_{s \star mb}$ is the solid angle of the convolution between the source size and the beam.
Usually, for a point source we can consider $\Omega_{s \star mb}\simeq\Omega_{mb}$. The $K/Jy$ conversion factors under this approximation are given in column 5 of Table~\ref{tab.caldetails}.
However, in order to account for the extended emission of Arp\,220 we assumed a source size $\theta_s=2''$,
based on the CO $J=2-1$ maps by \citet{Scoville1997}.
We can therefore correct for the source beam filling factor to calculate the source brightness temperature as
$T_{\rm b}=T_{\rm mb}\times\Omega_{s \star mb}/\Omega_{s}$ where $\Omega_{s}$ is the source solid angle.
If we use this in Eq.~\ref{eq.mbtojy} we obtain
\begin{equation}
\label{eq.btojy}
T_{\rm b}/S =\frac{\lambda^2}{2 k} \Omega_{s}^{-1}
\end{equation}
%where the source beam filling factor has been taken into account. 
Column 6 in Table~\ref{tab.caldetails} presents this conversion factor at
%into source brightness temperature for
each of the frequencies covered in this work.
We note that this conversion factors assume all molecules coexist and therefore no variations in the extent of the emission among molecules is considered.

Table~\ref{tab.NTrot} provides a compilation of the source averaged physical parameters, namely column densities ($N$) and rotational temperatures ($T_{\rm rot}$),
% and fractional abundances with respect to H$_2$, 
for each detected species and both velocity components.
$\rm H_2$ column density has been derived from the C$^{18}$O column density assuming a H$_2$/CO ratio  of $10^{-4}$ and the isotopical
$^{16}$O/$^{18}$O abundance ratio of 150, as derived for the starburst NGC\,253 \citep{Harrison1999}.

\subsubsection{Details on the fitting of individual species}
\label{Sect.FitDetails}
Here we describe the most relevant details on the fitting of particular species.
% following we describe the details of the observed $^{12}$CO $J=2-1$ emission and provide some details on the fitting of
% particular species.

% NS: If 200 K might account for 2.9 K km/s at 209 GHz, which would be <10% of the HC3N 23-22

$CH_3OH-$ The main methanol feature in the observed band is the $5-4$ group at $\sim241.8$\,GHz. This group of transitions together with a number of
transitions around 205.7 and 213.4\,GHz allow us to derive rotational temperatures of $T_{\rm rot}=15-20$\,K.
With this temperature we estimate the contribution of all the transitions present in the band.
It is however difficult to assess the presence of high temperature methanol due to the blending of most of its higher transitions.

$CH_3CCH-$ With the $J=12-11$ and $J=13-12$ groups of transitions of methyl acetylene we derive rotational temperatures of $T_{\rm rot}=16-21$\,K,
similar to those from methanol. With these parameters we have estimated the contribution of the $J=14-13$ group to the emission
of the CH$_3$CN spectral feature at 239\,GHz.

$SO-$ Only the $5_4-4_3$ and $5_5-4_4$ transitions at 206.1 and 215.2\,GHz are detected unambiguously and not
affected by blending to other lines.
Both have similar upper energy levels and therefore result in an unreliable measurement of the excitation temperature.
The parameters derived from these transitions were applied to estimate the $5_6-4_5$ line emission at 219.9\,GHz, observed between the $\rm ^{13}CO$ and $\rm C^{18}O$ lines.
The estimated contribution of this transition is not very sensitive to the accuracy of the determined rotational temperature.

$CH_3CN-$ together with HC$_3$N, CH$_3$CN is the only species in the survey showing clear evidence of highly excited
states. We have detected the emission from CH$_3$CN in both the ground and the $v_8=1$ vibrationally excited states.
We fitted simultaneously the $J=11-10$ and $J=13-12$ $v_8=0,1$ groups of transitions, as the $J=12-11$ is partially blended with $^{13}$CO.
The derived excitation temperatures are $\sim400-450$\,K.
Although the uncertainty in the excitation temperature is large ($20-30\%$), the high energy vibrational transitions are confirmed
in the three observed groups of CH$_3$CN transitions.
The estimated intensities of the $J=12-11$ group agree well with the observed line profile intensities and also show clear contribution of
the high excitation transitions.
The groups of transitions at 202.8, 221.3, and 239.7\,GHz, identified in Fig.~\ref{fig.specKmodel}, are not shown in Table~\ref{tab.ModelFit},
given that these features are due to the contribution of a few dozens of transitions with expected peak intensities at a 
%$\frac{1}{10}\sigma$ 
$0.1\sigma$ level. 

$H_2CO-$ We find only one line of H$_2$CO unblended, namely the $3_{1,2}-2_{1,1}$ transition at 225.7\,GHz.
The $3_{1,3}-2_{1,2}$ transition at 211.2\,GHz appears to be only partially blended with the faint $^{34}$SO and the vibrational HC$_3$N emission.
However, these transitions do not have enough dynamic range in their upper energy level to get an accurate determination of the excitation temperature.
From the fit to these transitions, the contribution of the H$_2$CO transition at 218\,GHz to the HC$_3$N $J=24-23$ observed line profiles was estimated.
% The fit is not very good. Over and Under estimated respectively.

$c$-$C_3H_2-$ Although no accurate temperature can be derived from the multiple transitions of c-C$_3$H$_2$, a better fit to the observed spectrum is
obtained with a high excitation temperature of $\sim 100$\,K. The uncertainty in the column density derived is less than a factor of two compared to
that from the fit obtained for a $T_{\rm rot}=20$\,K.

$HC_3N-$ Vibrationally excited emission of $\rm HC_3N$ towards Arp\,220 has been reported by (Mart\'in-Pintado, in prep).
Their detection of HC$_3$N in the $v_7=1$ and, tentatively, $v_6=1$ vibrational states are confirmed in all the four
groups of HC$_3$N transitions covered in this survey.
Moreover, we clearly detect the contribution of the emission of HC$_3$N in the $v_7=2$ state.
In order to avoid an excessive bias towards very high temperatures in the fitting, we used only the pure rotational transition and the $v_7=1$, $v_6=1$, and $v_7=2$
vibrational states.
Subsequently, given the high temperatures resulting from the fit, the significant contribution of $v_4=1$ and $v_5=1/v_7=1$ lines was calculated from the derived parameters.
The fitting to this molecule was performed individually for each group of ro-vibrational transitions, as the model of all the spectral features of HC$_3$N in the survey
did not accurately fit the observations.
Thus individual measurements of the excitation temperature have been derived for each group.
The derived temperatures averaged for both velocity components are  $\sim 320$, 380, and 440\,K for the $J_{upper}=23$, 25, and
26 ro-vibrational groups, respectively.
The $J=24-23$ group of ro-vibrational transitions are significantly blended with the $\rm C^{18}O$ and $\rm H_2CO$ transitions.
The excitation temperature was fixed to 355\,K as estimated for the $J_{upper}=24$ transition from the temperatures derived from the other groups.

$HNCO-$ No unblended transition of HNCO is detected so no additional information on this species can be derived from this
survey. The two brightest HNCO features are blended with C$^{18}$O and CH$_3$OH.
To account for the contribution of HNCO to the model spectra we considered its column density to be a factor of 200 with respect to C$^{18}$O
\citep{Mart'in2009}.

$C^{18}O-$ The $J=2-1$ transition of $\rm C^{18}O$ appears in one of the most crowded regions of the whole spectrum.
Its emission is blended with transitions of SO, HNCO, CH$_3$OH, and vibrationally excited states of HC$_3$N.
We fitted both HNCO and C$^{18}$O simultaneously, with their column density ratio fixed \citep[200:1,][]{Mart'in2009},
to the resulting spectrum after having subtracted the estimated contribution of SO and HC$_3$N.

$^{13}CO-$ The $J=2-1$ transition of $\rm ^{13}CO$ is partially blended with the CH$_3$CN $J=12-11$ group of transitions.
The transition of $\rm ^{13}CO$ was fitted after subtracting the estimated CH$_3$CN contribution.

$H_2^{18}O-$ We tentatively identify the spectral feature at 203.4\,GHz as the $3_{1,3}-2_{2,0}$ transition of $\rm H_2^{18}O$.
Details on the certainty of the identification are given in Sect.~\ref{sect.watermaser}.
This is the only species where a single Gaussian fit was performed in order to compare with the published results from the main isotopologue, H$_2$O \citep{Cernicharo2006}.
Fit results are shown in Table~\ref{tab.Water}.

$H^{13}CC_2N-$ We detected the emission of the three $^{13}$C isotopologues of HC$_3$N.
Both HC$^{13}$CCN and HCC$^{13}$CN emit at the same frequencies, so we assumed both species to have the same abundance.
At least one feature of these two isotopologues appears not to be blended to any other identified species.
H$^{13}$CCCN could not be independently fitted due to the blending of the transitions so a similar abundance to the other isotopologues was assumed.
For the three isotopologues we used a rotational temperature of 35\,K similar to the one derived from the pure rotational transitions of HC$_3$N
(see Sect.~\ref{sect.highTemp}).

$^{29}SiO-$ The spectral feature at 214.3\,GHz is tentatively identified as the $^{29}$Si isotopical substitution of SiO.
This feature could have a significant contribution from $\rm ^{13}CH_3CN$.
The $^{30}$SiO transition at 211.8\,GHz lies in between two observed frequency setups
and therefore we could not identify this transition which was expected to be a factor of 1.5 fainter than that
of $^{29}$SiO \citep{Penzias1981}.
%so the identification of a transition expected to be 1.5 fainter than $^{29}$SiO
%\citep{Penzias1981} is not feasible.

$^{13}CH_3CN-$ The detection of this isotopolgue is unclear and therefore is not indicated in Fig.~\ref{fig.specKmodel} or in the census presented
in Table~\ref{tab:census}.
However, from the two groups of transitions at $\sim 214$ and 230\,GHz, the latter in the edge of a band, we can estimate an upper limit to
its abundance.
We used the same rotational temperatures derived from CH$_3$CN for this estimate.
The feature at 214.3\,GHz cannot be accounted for by the emission of this isotopologue.
Moreover, a lower temperature would change the absolute column density derived but does not
result in a different contribution to the 214.3\,GHz line.
This limit supports the identification of $^{29}$SiO at this frequency.

$CH_2CO-$ The emission of $\rm CH_2CO$ could explain part of the spectral feature centered at 224.3\,GHz and the residual emission
observed not to be fully fitted by CH$_3$CCH at $\sim 222$\,GHz.
We fixed the excitation temperature to 20\,K and fitted the rest of lines to get an estimate of the column density. Then the contribution to the 
222 and 224.3\,GHz features was calculated. The contribution at 222\,GHz perfectly reproduces the observed feature, which supports its identification.

\subsubsection{Recombination lines}
Two hydrogen recombination lines are covered by our data at rest frequencies of 210.5\,GHz (H31$\alpha$) and 231.9\,GHz (H30$\alpha$).

A detection of H31$\alpha$ towards Arp\,220 was reported by \citet{Anantharamaiah2000} with a peak flux density of $\sim60$~mJy.
The width of their reported transition is much wider than the transitions they observed at radio frequencies.
Their single dish line profile agrees well with the line strength and shape of the emission we observe at 210.5\,GHz.
However, we identify this emission as vibrational transitions of HC$_3$N, which we confirm at other frequencies.
Indeed, \citet{Anantharamaiah2000} found their observed H31$\alpha$ line to be an order of magnitude brighter than predicted by their models.
Both the line profile and the confirmed emission of vibrationally excited HC$_3$N emission makes it seem that it was not H31$\alpha$ that they detected.
%cast doubts on the detection of H31$\alpha$.
Line contamination or misidentifications might also be considered for the other mm transitions they report at 3mm (H40$\alpha$ and H42$\alpha$),
where they also observed the emission to be much brighter and broader than expected.
This is not surprising given that the abundance of molecular transitions reported in this work was not known at the time.
Our results yield that the mm recombination lines towards Arp\,220 are much fainter that reported, which is
in better agreement with the single ionized component model by \citet{Anantharamaiah2000}

H30$\alpha$ lies right in the middle point between two observing setups, in one of the 127\kms\, gaps in the survey.
Should this line be bright enough, given the broad emission of the recombination lines in Arp\,220 of $\sim350$\kms\, \citep{Anantharamaiah2000},
we might expect to detect the edges of the line.
However, no signs of emission are observed in any of the neighboring frequencies which implies a non prominent H30$\alpha$ emission as
predicted by the the models from \citet{Anantharamaiah2000}.

\subsubsection{Unidentified features}
The LTE approximation appears to be good enough to fit most of the spectral features in the 40\,GHz surveyed in this study.
%frequency band.
However there are a number of features which are not accurately fitted.
There are several significant features which are not yet explained
such as the residual emission from the CH$_3$CN fit at $\sim221$\,GHz or that around CN at $\sim226$\,GHz and $\sim227$\,GHz.
Similarly the emission at $\sim229.5$\,GHz is not fitted by the vibrational states of HC$_3$N.
Additionally, we observe significant residual emission between C$^{34}$S and methanol at $\sim241.5$\,GHz.
All these residual emission features could be due either to contribution from unidentified molecules, or non-LTE emission from the identified species,
or a departure from the kinematical constraints on the two Gaussian components applied in the modeling.

On the other hand, there are other unblended features that must be due to unidentified molecular species or transitions under non-LTE conditions.
Such is the case of the feature at $\sim 212.1$\,GHz in the edge of the 213\,GHz frequency setup.
With a peak intensity of $\sim 150$\,mK,
at a $\sim7\sigma$ level above the rms noise, this is likely an unidentified molecular species.
Even more significant is the double 200\,mK peak features at 224.0 and 224.4\,GHz.
The emission from the tentatively detected CH$_2$CO can partially account
for this feature but fails to reproduce the broad double-peak emission profile.
Finally the broad feature at 234.3\,GHz, though tentative, 
%might also be the trace of a unidentified transition but uncertainty in the relative continuum
%calibration cannot be discarded.
might be from an unidentified line transition although we can not rule out the possibility of a residual in the continuum subtraction.

Exploration of a broader band covering more transitions and deeper integration would be necessary to accurately identify these features.

\begin{table}
\begin{center}
\scriptsize
\caption{Fitted parameters to the identified transitions}
\label{tab.ModelFit}
\begin{tabular}{c @{\,\,\,} l @{} r @{\,\,\,} r @{\,\,\,} r @{\,\,\,} r}
\hline
\hline
                &                                        & \multicolumn{2}{c}{Component 1}               &    \multicolumn{2}{c}{Component 2}            \\
$\nu$           & Molecule                               & $\int{T_{\rm B}{\rm d}v}$   & $T_{\rm B}$     &  $\int{T_{\rm B}{\rm d}v}$   &   $T_{\rm B}$  \\
(GHz)           & Transition                             & K\,km\,s$^{-1}$             &   mK            &  K\,km\,s$^{-1}$             &    mK          \\
\hline
230.538         & $\rm CO\,\,2-1$                        &      3140                   &   11600         &   1820                       &    6700        \\
207.436         & $\rm NS\,\,9/2_{11/2}-7/2_{9/2}\,e$    &      17.0                   &   62.8          &  24.7                        &  91.2          \\
207.436         & $\rm NS\,\,9/2_{9/2}-7/2_{7/2}\,e$     &      13.5                   &   49.7          &  19.6                        &  72.3          \\
207.438         & $\rm NS\,\,9/2_{7/2}-7/2_{5/2}\,e$     &      10.6                   &   39.2          &  15.4                        &  57.0          \\
207.834         & $\rm NS\,\,9/2_{11/2}-7/2_{9/2}\,f$    &      17.0                   &   62.6          &  24.6                        &  91.0          \\
207.838         & $\rm NS\,\,9/2_{9/2}-7/2_{7/2}\,f$     &      13.4                   &   49.6          &  19.5                        &  72.0          \\
207.838         & $\rm NS\,\,9/2_{7/2}-7/2_{5/2}\,f$     &      10.6                   &   39.1          &  15.4                        &  56.9          \\
241.016         & $\rm C^{34}S\,\,5-4$                   &     100.2                   &   369.5         &  107.5                       &  396.5         \\
231.221         & $\rm ^{13}CS\,\,5-4$                   &      31.2                   &   115.1         &   72.5                       &  267.4         \\
235.380         & $\rm CO^+\,\,2_{3/2}-1_{3/2}$          &      ...                    &    ...          &    9.3                       &  34.4          \\
235.791         & $\rm CO^+\,\,2_{3/2}-1_{1/2}$          &      14.6                   &    53.8         &    46.7                      &  172.2         \\
236.062         & $\rm CO^+\,\,2_{5/2}-1_{3/2}$          &      26.2                   &    96.8         &    84.0                      &  309.8         \\
205.791         & $\rm CH_3OH\,\,1_{1,1}-2_{0,2}\, A+ $  &      25.7                   &    94.8         &    39.3                      &  145.1         \\
213.427         & $\rm CH_3OH\,\,1_{1,0}-0_{0,0}\, E $   &      15.7                   &    57.9         &    26.4                      &   97.3         \\
218.440         & $\rm CH_3OH\,\,4_{2,2}-3_{1,2}\, E $   &      14.8                   &    54.7         &    34.3                      &  126.7         \\
230.027         & $\rm CH_3OH\,\,3_{-2,2}-4_{-1,4}\, E $ &      ...                    &    ...          &   10.2                       &   37.6         \\
239.746         & $\rm CH_3OH\,\,5_{1,5}-4_{1,4}\, A+ $  &      14.4                   &    53.3         &   35.2                       &  130.0         \\
241.700         & $\rm CH_3OH\,\,5_{0,5}-4_{0,4}\, E $   &      16.3                   &    60.1         &   39.1                       &  144.2         \\
241.767         & $\rm CH_3OH\,\,5_{-1,5}-4_{-1,4}\, E $ &      25.6                   &    94.5         &   55.1                       &  203.2         \\
241.791         & $\rm CH_3OH\,\,5_{0,5}-4_{0,4}\, A+ $  &      38.4                   &   141.6         &   76.1                       &  280.7         \\
241.842         & $\rm CH_3OH\,\,5_{2,4}-4_{2,3}\, A- $  &      ...                    &    ...          &    9.5                       &   35.1         \\
241.879         & $\rm CH_3OH\,\,5_{1,4}-4_{1,3}\, E $   &      9.6                    &    35.5         &   25.8                       &   95.1         \\
241.887         & $\rm CH_3OH\,\,5_{2,3}-4_{2,2}\, A+ $  &      ...                    &    ...          &    9.5                       &   35.1         \\
241.904         & $\rm CH_3OH\,\,5_{-2,4}-4_{-2,3}\, E $ &      ...                    &    ...          &   17.2                       &   63.5         \\
241.905         & $\rm CH_3OH\,\,5_{2,3}-4_{2,2}\, E $   &      7.5                    &    27.6         &   20.5                       &   75.5         \\
205.065         & $\rm CH_3CCH\,\,12_2-11_2 $            &     ...                     &    ...          &    6.9                       &   25.4         \\
205.077         & $\rm CH_3CCH\,\,12_1-11_1 $            &     17.2                    &    63.4         &   19.2                       &   70.8         \\
205.081         & $\rm CH_3CCH\,\,12_0-11_0 $            &     27.2                    &   100.2         &   27.0                       &   99.7         \\
222.163         & $\rm CH_3CCH\,\,13_1-12_1 $            &     10.4                    &    38.4         &   13.8                       &   50.8         \\
222.167         & $\rm CH_3CCH\,\,13_0-12_0 $            &     16.4                    &    60.7         &   19.4                       &   71.5         \\
239.248         & $\rm CH_3CCH\,\,14_1-13_1 $            &     ...                     &    ...          &    9.4                       &   34.6         \\
239.252         & $\rm CH_3CCH\,\,14_0-13_0 $            &      9.3                    &    34.4         &   13.2                       &   48.7         \\
206.176         & $\rm SO\,\,5_4-4_3 $                   &     31.9                    &   117.6         &   88.1                       &  325.1         \\
215.221         & $\rm SO\,\,5_5-4_4 $                   &     30.1                    &   111.1         &   91.6                       &  337.8         \\
219.949         & $\rm SO\,\,5_6-4_5 $                   &     67.0                    &   247.1         &  173.8                       &  641.0         \\
217.105         & $\rm SiO\,\,5-4 $                      &     75.9                    &   280.1         &  128.3                       &  473.4         \\
216.710         & $\rm H_2S\,\,2_2-2_1 $                 &    115.8                    &   427.1         &  107.6                       &  396.8         \\
211.211         & $\rm H_2CO\,\,3_{1,3}-2_{1,2} $        &     61.0                    &   225.2         &   93.4                       &  344.4         \\
218.222         & $\rm H_2CO\,\,3_{0,3}-2_{0,2} $        &     51.0                    &   188.4         &   73.9                       &  272.7         \\
225.698         & $\rm H_2CO\,\,3_{1,2}-2_{1,1} $        &     59.5                    &   219.4         &   91.6                       &  337.8         \\
213.624         & c-$\rm C_3H_2\,\,14_{9,6}-14_{8,7} $   &      6.9                    &    25.5         &   ...                        &    ...         \\
214.621         & c-$\rm C_3H_2\,\,13_{7,6}-13_{6,7} $   &      9.0                    &    33.4         &   ...                        &    ...         \\
215.974         & c-$\rm C_3H_2\,\,12_{7,6}-12_{6,7} $   &     11.3                    &    41.7         &   ...                        &    ...         \\
216.279         & c-$\rm C_3H_2\,\, 3_{3,0}- 2_{2,1} $   &     17.4                    &    64.1         &   15.9                       &   58.6         \\
216.809         & c-$\rm C_3H_2\,\,11_{5,6}-11_{4,7} $   &     13.4                    &    49.5         &    7.6                       &   28.1         \\
217.532         & c-$\rm C_3H_2\,\,10_{5,6}-10_{4,7} $   &     14.9                    &    55.0         &    9.4                       &   34.6         \\
217.822         & c-$\rm C_3H_2\,\, 6_{1,6}- 5_{0,5} $   &     57.0                    &   210.3         &   49.3                       &  181.8         \\
217.822         & c-$\rm C_3H_2\,\, 6_{0,6}- 5_{1,5} $   &     19.0                    &    70.1         &   16.4                       &   60.6         \\
217.940         & c-$\rm C_3H_2\,\, 5_{1,4}- 4_{2,3} $   &     37.0                    &   136.3         &   32.2                       &  118.9         \\
218.055         & c-$\rm C_3H_2\,\, 9_{3,6}- 9_{2,7} $   &     15.2                    &    56.2         &   10.5                       &   38.8         \\
218.160         & c-$\rm C_3H_2\,\, 5_{2,4}- 4_{1,3} $   &     12.3                    &    45.5         &   10.8                       &   39.7         \\
218.449         & c-$\rm C_3H_2\,\, 8_{3,6}- 8_{2,7} $   &     13.6                    &    50.0         &   10.2                       &   37.6         \\
218.733         & c-$\rm C_3H_2\,\, 7_{1,6}- 7_{0,7} $   &      8.9                    &    32.8         &    7.2                       &   26.6         \\
227.169         & c-$\rm C_3H_2\,\, 4_{3,2}- 3_{2,1} $   &     22.7                    &    83.8         &   20.2                       &   74.5         \\
209.230         & $\rm HC_3N\,\,23-22 $                  &  231.5                      &  854.2          & 241.6                        &  891.2         \\
209.744         & $\rm HC_3N\,\,23-22\,v_7=1_{l=1e} $    &   89.1                      &  328.5          &  83.1                        &  306.5         \\
210.044         & $\rm HC_3N\,\,23-22\,v_7=1_{l=1f} $    &   89.1                      &  328.8          &  83.2                        &  306.8         \\
210.527         & $\rm HC_3N\,\,23-22\,v_7=2_{l=0}  $    &   34.4                      &  127.0          &  28.7                        &  105.9         \\
210.555         & $\rm HC_3N\,\,23-22\,v_7=2_{l=2e} $    &   33.8                      &  124.8          &  28.2                        &  104.0         \\
210.586         & $\rm HC_3N\,\,23-22\,v_7=2_{l=2f} $    &   33.8                      &  124.8          &  28.2                        &  104.0         \\
209.573         & $\rm HC_3N\,\,23-22\,v_6=1_{l=1e} $    &   27.5                      &  101.6          &  22.4                        &   82.5         \\
209.738         & $\rm HC_3N\,\,23-22\,v_6=1_{l=1f} $    &   27.6                      &  101.6          &  22.4                        &   82.6         \\
209.245         & $\rm HC_3N\,\,23-22\,v_5=1/v_7=3  $    &   13.5                      &   49.8          &  10.1                        &   37.3         \\
209.362         & $\rm HC_3N\,\,23-22\,v_5=1/v_7=3  $    &   13.5                      &   49.8          &  10.1                        &   37.3         \\
210.904         & $\rm HC_3N\,\,23-22\,v_5=1/v_7=3  $    &   13.5                      &   50.0          &  10.1                        &   37.4         \\
211.225         & $\rm HC_3N\,\,23-22\,v_5=1/v_7=3  $    &   13.1                      &   48.3          &   9.8                        &   36.0         \\
211.226         & $\rm HC_3N\,\,23-22\,v_5=1/v_7=3  $    &   13.1                      &   48.3          &   9.8                        &   36.0         \\
211.507         & $\rm HC_3N\,\,23-22\,v_5=1/v_7=3  $    &   13.6                      &   50.0          &  10.1                        &   37.4         \\
218.325         & $\rm HC_3N\,\,24-23 $                  &  238.7                      &  880.4          & 141.8                        &  523.2         \\
218.861         & $\rm HC_3N\,\,24-23\,v_7=1_{l=1e} $    &   96.2                      &  354.7          &  57.2                        &  210.8         \\
219.174         & $\rm HC_3N\,\,24-23\,v_7=1_{l=1f} $    &   96.3                      &  355.1          &  57.2                        &  211.0         \\
219.675         & $\rm HC_3N\,\,24-23\,v_7=2_{l=0}  $    &   38.9                      &  143.6          &  23.2                        &   85.4         \\
219.707         & $\rm HC_3N\,\,24-23\,v_7=2_{l=2e} $    &   38.3                      &  141.3          &  22.8                        &   84.0         \\
219.742         & $\rm HC_3N\,\,24-23\,v_7=2_{l=2f} $    &   38.3                      &  141.3          &  22.8                        &   84.0         \\
218.682         & $\rm HC_3N\,\,24-23\,v_6=1_{l=1e} $    &   31.5                      &  116.1          &  18.7                        &   69.0         \\
218.854         & $\rm HC_3N\,\,24-23\,v_6=1_{l=1f} $    &   31.5                      &  116.2          &  18.7                        &   69.1         \\
218.340         & $\rm HC_3N\,\,24-23\,v_5=1/v_7=3  $    &   16.0                      &   58.9          &   9.5                        &   35.0         \\
218.462         & $\rm HC_3N\,\,24-23\,v_5=1/v_7=3  $    &   16.0                      &   58.9          &   9.5                        &   35.0         \\
220.070         & $\rm HC_3N\,\,24-23\,v_5=1/v_7=3  $    &   16.0                      &   59.1          &   9.5                        &   35.1         \\
220.407         & $\rm HC_3N\,\,24-23\,v_5=1/v_7=3  $    &   15.5                      &   57.3          &   9.2                        &   34.0         \\
220.408         & $\rm HC_3N\,\,24-23\,v_5=1/v_7=3  $    &   15.5                      &   57.3          &   9.2                        &   34.0         \\
220.700         & $\rm HC_3N\,\,24-23\,v_5=1/v_7=3  $    &   16.0                      &   59.2          &   9.5                        &   35.2         \\
\hline
\end{tabular}
\end{center}
\end{table}

\begin{table}
\begin{center}
\scriptsize
\addtocounter{table}{-1}
\caption{(Cont.)}
\begin{tabular}{c @{\,\,\,} l @{} r @{\,\,\,} r @{\,\,\,} r @{\,\,\,} r}
\hline
\hline
                &                                        & \multicolumn{2}{c}{Component 1}               &    \multicolumn{2}{c}{Component 2}            \\
$\nu$           & Molecule                               & $\int{T_{\rm B}{\rm d}v}$   & $T_{\rm B}$     &  $\int{T_{\rm B}{\rm d}v}$   &   $T_{\rm B}$  \\
(GHz)           & Transition                             & K\,km\,s$^{-1}$             &   mK            &  K\,km\,s$^{-1}$             &    mK          \\
\hline
227.419         & $\rm HC_3N\,\,25-24 $                  &  175.2                      &  646.4          & 129.9                        &  479.1         \\
227.977         & $\rm HC_3N\,\,25-24\,v_7=1_{l=1e} $    &   71.5                      &  263.7          &  59.0                        &  217.5         \\
228.303         & $\rm HC_3N\,\,25-24\,v_7=1_{l=1f} $    &   71.5                      &  263.9          &  59.0                        &  217.6         \\
228.822         & $\rm HC_3N\,\,25-24\,v_7=2_{l=0}  $    &   29.3                      &  108.0          &  26.9                        &   99.1         \\
228.859         & $\rm HC_3N\,\,25-24\,v_7=2_{l=2e} $    &   28.8                      &  106.4          &  26.5                        &   97.7         \\
228.898         & $\rm HC_3N\,\,25-24\,v_7=2_{l=2f} $    &   28.8                      &  106.4          &  26.5                        &   97.7         \\
227.792         & $\rm HC_3N\,\,25-24\,v_6=1_{l=1e} $    &   23.8                      &   87.6          &  22.4                        &   82.5         \\
227.971         & $\rm HC_3N\,\,25-24\,v_6=1_{l=1f} $    &   23.8                      &   87.7          &  22.4                        &   82.5         \\
227.435         & $\rm HC_3N\,\,25-24\,v_5=1/v_7=3  $    &   12.2                      &   44.9          &  12.4                        &   45.7         \\
227.562         & $\rm HC_3N\,\,25-24\,v_5=1/v_7=3  $    &   12.2                      &   44.9          &  12.4                        &   45.7         \\
229.236         & $\rm HC_3N\,\,25-24\,v_5=1/v_7=3  $    &   12.2                      &   45.0          &  12.4                        &   45.8         \\
229.589         & $\rm HC_3N\,\,25-24\,v_5=1/v_7=3  $    &   11.8                      &   43.7          &  12.1                        &   44.6         \\
229.590         & $\rm HC_3N\,\,25-24\,v_5=1/v_7=3  $    &   11.8                      &   43.7          &  12.1                        &   44.6         \\
229.891         & $\rm HC_3N\,\,25-24\,v_5=1/v_7=3  $    &   12.2                      &   45.1          &  12.4                        &   45.9         \\
236.513         & $\rm HC_3N\,\,26-25 $                  &  123.7                      &  456.3          & 115.8                        &  427.0         \\
237.093         & $\rm HC_3N\,\,26-25\,v_7=1_{l=1e} $    &   55.5                      &  204.8          &  59.0                        &  217.8         \\
237.432         & $\rm HC_3N\,\,26-25\,v_7=1_{l=1f} $    &   55.6                      &  205.0          &  59.1                        &  218.1         \\
237.969         & $\rm HC_3N\,\,26-25\,v_7=2_{l=0}  $    &   25.0                      &   92.3          &  30.2                        &  111.6         \\
238.010         & $\rm HC_3N\,\,26-25\,v_7=2_{l=2e} $    &   24.7                      &   91.0          &  29.9                        &  110.2         \\
238.054         & $\rm HC_3N\,\,26-25\,v_7=2_{l=2f} $    &   24.7                      &   91.0          &  29.9                        &  110.2         \\
236.900         & $\rm HC_3N\,\,26-25\,v_6=1_{l=1e} $    &   20.8                      &   76.6          &  25.9                        &   95.4         \\
237.086         & $\rm HC_3N\,\,26-25\,v_6=1_{l=1f} $    &   20.8                      &   76.6          &  25.9                        &   95.5         \\
236.184         & $\rm HC_3N\,\,26-25\,v_4=1        $    &   ...                       &   ...           &   8.1                        &   29.9         \\
236.529         & $\rm HC_3N\,\,26-25\,v_5=1/v_7=3  $    &   11.4                      &   42.1          &  15.6                        &   57.6         \\
236.661         & $\rm HC_3N\,\,26-25\,v_5=1/v_7=3  $    &   11.4                      &   42.1          &  15.6                        &   57.6         \\
238.401         & $\rm HC_3N\,\,26-25\,v_5=1/v_7=3  $    &   11.4                      &   42.2          &  15.7                        &   57.8         \\
238.770         & $\rm HC_3N\,\,26-25\,v_5=1/v_7=3  $    &   11.1                      &   41.1          &  15.3                        &   56.4         \\
238.772         & $\rm HC_3N\,\,26-25\,v_5=1/v_7=3  $    &   11.1                      &   41.1          &  15.3                        &   56.4         \\
239.082         & $\rm HC_3N\,\,26-25\,v_5=1/v_7=3  $    &   11.4                      &   42.3          &  15.7                        &   57.9         \\
219.798         & $\rm HNCO\,\,10_{10,11}-9_{9,10}    $  &   7.2                       &   26.5          &  11.8                        &   43.4        \\
219.798         & $\rm HNCO\,\,10_{10,10}-9_{9,9}     $  &   ...                       &   ...           &  10.6                        &   39.2        \\
219.798         & $\rm HNCO\,\,10_{10,9} -9_{9,8}     $  &   ...                       &   ...           &   9.6                        &   35.5        \\
241.774         & $\rm HNCO\,\,11_{11,12}-10_{10,11}  $  &   ...                       &   ...           &   7.9                        &   29.3        \\
241.774         & $\rm HNCO\,\,11_{11,11}-10_{10,10}  $  &   ...                       &   ...           &   7.2                        &   26.7        \\
219.560         & $\rm C^{18}O\,\,2-1                 $  &  185.0                      &  682.5          & 302.5                        & 1115.8        \\
220.399         & $\rm ^{13}CO\,\,2-1                 $  &  240.6                      &  887.7          & 273.3                        & 1008.3        \\
211.014         & $\rm ^{34}SO\,\,5_5-4_4 $              &  ...                        &   ...           &  32.5                        &  120.0        \\
215.839         & $\rm ^{34}SO\,\,5_6-4_5 $              &   13.5                      &   49.9          &  60.8                        &  224.2        \\
202.320         & $\rm CH_3CN\,\,11_3-10_3  $            &  12.5                        &   46.1           &  15.1                        &   55.7        \\
%202.320         & $\rm CH_3CN\,\,11_3-10_3  $            &  ...                        &   ...           &   7.2                        &   26.6        \\
%202.320         & $\rm CH_3CN\,\,11_3-10_3  $            &  ...                        &   ...           &   7.9                        &   29.1        \\
220.709         & $\rm CH_3CN\,\,12_3-11_3  $            &   21.1                       &   78.1          &  25.5                        &   94.3        \\
%220.709         & $\rm CH_3CN\,\,12_3-11_3  $            &   7.0                       &   26.0          &   8.5                        &   31.4        \\
%220.709         & $\rm CH_3CN\,\,12_3-11_3  $            &  ...                        &   ...           &   7.8                        &   28.8        \\
%220.709         & $\rm CH_3CN\,\,12_3-11_3  $            &   7.6                       &   28.2          &   9.2                        &   34.1        \\
239.096         & $\rm CH_3CN\,\,13_3-12_3  $            &  24.4                       &   89.9          &  29.5                        &  109.1        \\
%239.096         & $\rm CH_3CN\,\,13_3-12_3  $            &   8.1                       &   29.9          &   9.8                        &   36.3        \\
%239.096         & $\rm CH_3CN\,\,13_3-12_3  $            &   7.5                       &   27.7          &   9.1                        &   33.6        \\
%239.096         & $\rm CH_3CN\,\,13_3-12_3  $            &   8.8                       &   32.3          &  10.6                        &   39.2        \\
208.343         & $\rm HC^{13}CCN\,\,23-22  $            &  40.1                       &  148.0          &  21.2                        &   78.0        \\
217.398         & $\rm HC^{13}CCN\,\,24-23  $            &  32.5                       &  119.7          &  17.1                        &   63.1        \\
226.454         & $\rm HC^{13}CCN\,\,25-24  $            &  25.8                       &   95.3          &  13.6                        &   50.2        \\
235.509         & $\rm HC^{13}CCN\,\,26-25  $            &  20.2                       &   74.7          &  10.7                        &   39.4        \\
208.363         & $\rm HCC^{13}CN\,\,23_{22}-22_{21}  $  &  12.8                       &   47.1          &   ...                        &   ...         \\
208.363         & $\rm HCC^{13}CN\,\,23_{23}-22_{22}  $  &  13.3                       &   49.2          &   7.0                        &   26.0        \\
208.363         & $\rm HCC^{13}CN\,\,23_{24}-22_{23}  $  &  13.9                       &   51.4          &   7.3                        &   27.1        \\
217.420         & $\rm HCC^{13}CN\,\,24_{23}-23_{22}  $  &  10.4                       &   38.2          &   ...                        &   ...         \\
217.420         & $\rm HCC^{13}CN\,\,24_{24}-23_{23}  $  &  10.8                       &   39.8          &   ...                        &   ...         \\
217.420         & $\rm HCC^{13}CN\,\,24_{25}-23_{24}  $  &  11.3                       &   41.5          &   ...                        &   ...         \\
226.476         & $\rm HCC^{13}CN\,\,25-24  $            &  25.8                       &   95.3          &  13.6                        &   50.2        \\
235.532         & $\rm HCC^{13}CN\,\,26-25  $            &  20.2                       &   74.7          &  10.7                        &   39.4        \\
202.764         & $\rm H^{13}CCCN\,\,23-22  $            &  41.6                       &  153.6          &  22.0                        &   81.0        \\
211.577         & $\rm H^{13}CCCN\,\,24-23  $            &  34.0                       &  125.3          &  17.9                        &   66.0        \\
220.390         & $\rm H^{13}CCCN\,\,25-24  $            &  27.2                       &  100.6          &  14.4                        &   53.0        \\
229.203         & $\rm H^{13}CCCN\,\,26-25  $            &  21.5                       &  079.5          &  11.4                        &   41.9        \\
238.016         & $\rm H^{13}CCCN\,\,27-26  $            &  16.8                       &  061.9          &   8.8                        &   32.6        \\
214.385         & $\rm ^{29}SiO\,\,5-4  $                &  27.1                       &  100.1          &  39.4                        &  145.0        \\
202.014         & $\rm CH_2CO\,\,10_{0,10}- 9_{0, 9}  $  &   9.2                       &   34.0          &  25.7                        &   94.9        \\
203.940         & $\rm CH_2CO\,\,10_{1, 9}- 9_{1, 8}  $  &  14.0                       &   51.7          &  39.1                        &  144.5        \\
220.178         & $\rm CH_2CO\,\,11_{1,11}-10_{1,10}  $  &  10.4                       &   38.2          &  29.0                        &  106.8        \\
222.198         & $\rm CH_2CO\,\,11_{0,11}-10_{0,10}  $  &   ...                       &   ...           &  18.3                        &   67.7        \\
224.328         & $\rm CH_2CO\,\,11_{1,10}-10_{1, 9}  $  &  10.0                       &   36.7          &  27.8                        &  102.6        \\
240.187         & $\rm CH_2CO\,\,12_{1,12}-11_{1,11}  $  &   7.0                       &   25.7          &  19.5                        &   71.8        \\
226.632         & $\rm CN\,\,2_{3/2-3/2}-1_{1/2,3/2}  $  &   ...                       &   ...           &  11.6                        &   42.7        \\
226.660         & $\rm CN\,\,2_{3/2-5/2}-1_{1/2-3/2}  $  &  13.8                       &   50.7          &  38.6                        &  142.4        \\
226.664         & $\rm CN\,\,2_{3/2-1/2}-1_{1/2-1/2}  $  &   ...                       &   ...           &  11.5                        &   42.4        \\
226.679         & $\rm CN\,\,2_{3/2-3/2}-1_{1/2-1/2}  $  &   ...                       &   ...           &  14.3                        &   52.8        \\
226.874         & $\rm CN\,\,2_{5/2-5/2}-1_{3/2-3/2}  $  &  13.9                       &   51.4          &  39.1                        &  144.3        \\
226.875         & $\rm CN\,\,2_{5/2-7/2}-1_{3/2-5/2}  $  &  22.1                       &   81.4          &  62.0                        &  228.6        \\
226.876         & $\rm CN\,\,2_{5/2-3/2}-1_{3/2-1/2}  $  &   8.3                       &   30.6          &  23.3                        &   85.8        \\
226.887         & $\rm CN\,\,2_{5/2-3/2}-1_{3/2-3/2}  $  &   ...                       &   ...           &   7.4                        &   27.3        \\
226.892         & $\rm CN\,\,2_{5/2-5/2}-1_{3/2-5/2}  $  &   ...                       &   ...           &   7.4                        &   27.1        \\
\hline
\end{tabular}
\end{center}
\end{table}

\begin{table}
\begin{center}
\scriptsize
\caption{Derived LTE physical parameters for each velocity component \label{tab.NTrot}}
\begin{tabular}{l c @{$-$} c c @{$-$} c}
\hline
\hline
Molecule            &    \multicolumn{2}{c}{$N$}                 &   \multicolumn{2}{c}{$T_{\rm rot}$}         \\
                    &  \multicolumn{2}{c}{$10^{14}$(cm$^{-2}$)}  &   \multicolumn{2}{c}{(K)}                   \\
                    &      Comp.1  &  Comp.2 &  Comp.1 &  Comp.2  \\    
\hline
CO                  &      16565.8 & 9527.0  &   \multicolumn{2}{c}{...}          \\
$^{13}$CO           &       1327.7 & 1508.2  &   \multicolumn{2}{c}{...}          \\
C$^{18}$O           &       1024.7 & 1675.3  &   \multicolumn{2}{c}{...}          \\
H$_2$S              &        283.2 & 263.1   &   \multicolumn{2}{c}{...}           \\
CH$_3$CCH           &        130.2 & 70.0    &   16  &  21       \\
CH$_3$OH            &         50.4 & 89.7    &   15  &  20       \\
HC$_3$N\,\tablefootmark{a}     &         29.8 & 25.8    &   36  &  35       \\
HC$_3$N$^*$\,\tablefootmark{a} &         10.1 & 8.2     &  358  &  376      \\
CH$_2$CO            &          9.7 & 27.1    &   \multicolumn{2}{c}{...}          \\
SO                  &         10.1 & 21.4    &   16  &  22        \\
CH$_3$CN            &          7.5 & 12.1    &   392 &  469        \\
c-C$_3$H$_2$        &          9.8 & 6.1     &  117  &  87        \\
HNCO                &          5.1 & 8.3     &   \multicolumn{2}{c}{...}          \\
CO$^+$              &          2.7 & 8.7     &   \multicolumn{2}{c}{...}          \\
$^{34}$SO           &          2.2 & 7.5     &   15  &  24      \\
CN                  &          2.5 & 7.2     &   \multicolumn{2}{c}{...}          \\
NS                  &          3.4 & 5.0     &   \multicolumn{2}{c}{...}          \\
H$^{13}$CC$_2$N\,\tablefootmark{b}  & 5.0 & 2.7 &   \multicolumn{2}{c}{35 \tablefootmark{c}}          \\
H$_2$CO             &          3.0 & 4.4    &   14  & 16         \\
C$^{34}$S           &          3.6 & 3.9    &   \multicolumn{2}{c}{...}          \\
$^{13}$CS           &          1.1 & 2.7    &   \multicolumn{2}{c}{...}          \\
SiO                 &          1.1 & 1.9    &   \multicolumn{2}{c}{...}          \\
$^{29}$SiO          &          0.4 & 0.6    &   \multicolumn{2}{c}{...}          \\
$^{13}$CH$_3$CN     &       $<0.5$ & $<0.8$ &  390\tablefootmark{b} & 470\tablefootmark{c}         \\
\hline
\end{tabular}
\tablefoot{
\tablefoottext{a}{HC$_3$N parameters shown for the fitting to the ground vibrationally state transitions
alone (HC$_3$N) and the fit to all observed vibrational transitions (HC$_3$N$^*$).
See text in Sect.~\ref{Sect.FitDetails}}
\tablefoottext{b}{Refers to either H$^{13}$CCCN, HC$^{13}$CCN, or HCC$^{13}$CN.}
\tablefoottext{c}{These temperatures were fixed to match those derived in with the main isotopologues.}
}
\end{center}
\end{table}

\section{Discussion}

\subsection{Chemical composition of Arp\,220}
\begin{figure}
\resizebox{\hsize}{!}{\includegraphics[angle=-90,width=0.5\textwidth]{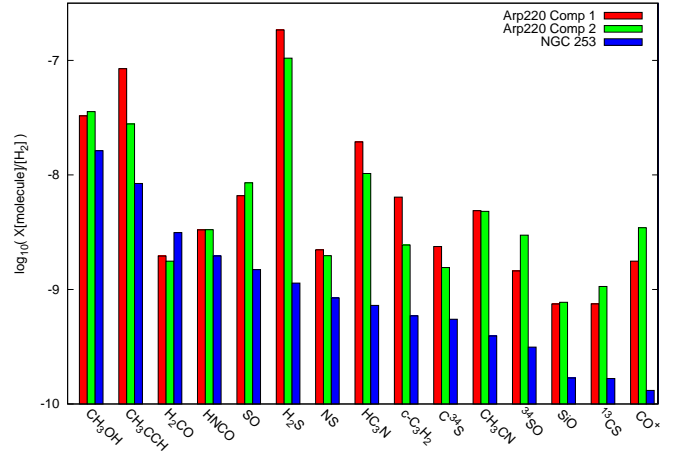}}
\caption{Comparison of the fractional abundances relative to H$_2$ derived for the two velocity components in Arp\,220 and those
measured in the starburst galaxy NGC\,253. Abundances for NGC\,253 derived \citet{Mart'in2006} and H$_2$ derived from C$^{18}$O \citep{Harrison1999}.
\label{fig.AbunComp}}
\end{figure}

The column densities of 18 molecular species and isotopologues other than CO derived from the line survey in this work allow us to carry out a detailed study
of chemical composition of the nuclear ISM in Arp\,220.
In Fig.~\ref{fig.AbunComp} we compare the derived fractional abundances relative to H$_2$ in Arp\,220 with those in the prototypical starburst
NGC\,253.
Molecular hydrogen column densities, as detailed in Sect.~\ref{sect.LTE}, were derived for both sources from C$^{18}$O assuming a $^{16}$O/$^{18}$O=150
as estimated for NGC\,253 \citep{Harrison1999}.
For Arp\,220 we derive $N_{\rm H_2}=1.5\times10^{23}\,\rm cm^{-2}$ and $2.5\times10^{23}\,\rm cm^{-2}$ for the velocity components 1 and 2, respectively.
The direct comparison between the chemical composition of both galaxies show clear overabundances of H$_2$S, HC$_3$N, CH$_3$CN, and CO$^+$ towards Arp\,220.
Moreover, there is an apparent general overabundance of all the detected molecular species in Arp\,220 except for H$_2$CO.
By averaging the measured abundances in both galaxies for all species but those mentioned with clear differences in their relative abundances, we
notice that the averaged molecular abundances in Arp\,220 are
%can estimate the dense gas ratio to the overall molecular gas to be 
a factor of $\sim4.5$ and $\sim3$ higher for the 5208 and 5462\,\kms\, velocity components with respect the averaged abundance in NGC\,253.
Such a difference might be related to the uncertainty in the estimate of the H$_2$ column density from C$^{18}$O in both galaxies.
The $^{16}$O/$^{18}$O ratio is subject to a large uncertainty in NGC\,253 \citep{Mart'in2010a} and is likely to be different in Arp\,220.
The discrepancy might be even larger if the overproduction of C$^{18}$O proposed based on CO isotopologues ratios \citep{Matsushita2009} is confirmed.
However, as discussed in Sect.~\ref{sect.isotRat}, this increase would be marked by a significant optical depth in C$^{18}$O.

\begin{figure}
\resizebox{\hsize}{!}{\includegraphics[angle=-90]{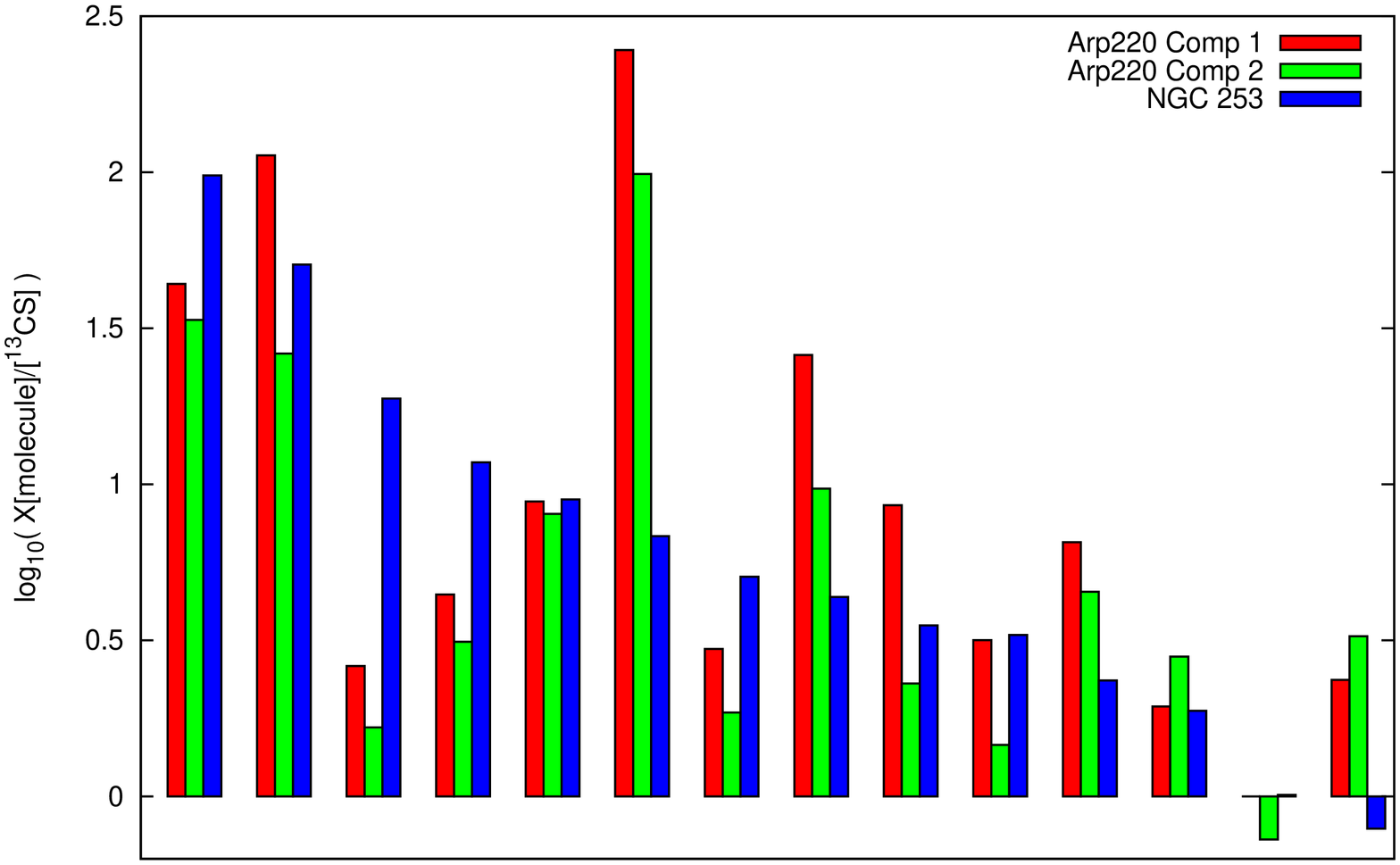}}\\[-13pt]
\resizebox{\hsize}{!}{\includegraphics[angle=-90]{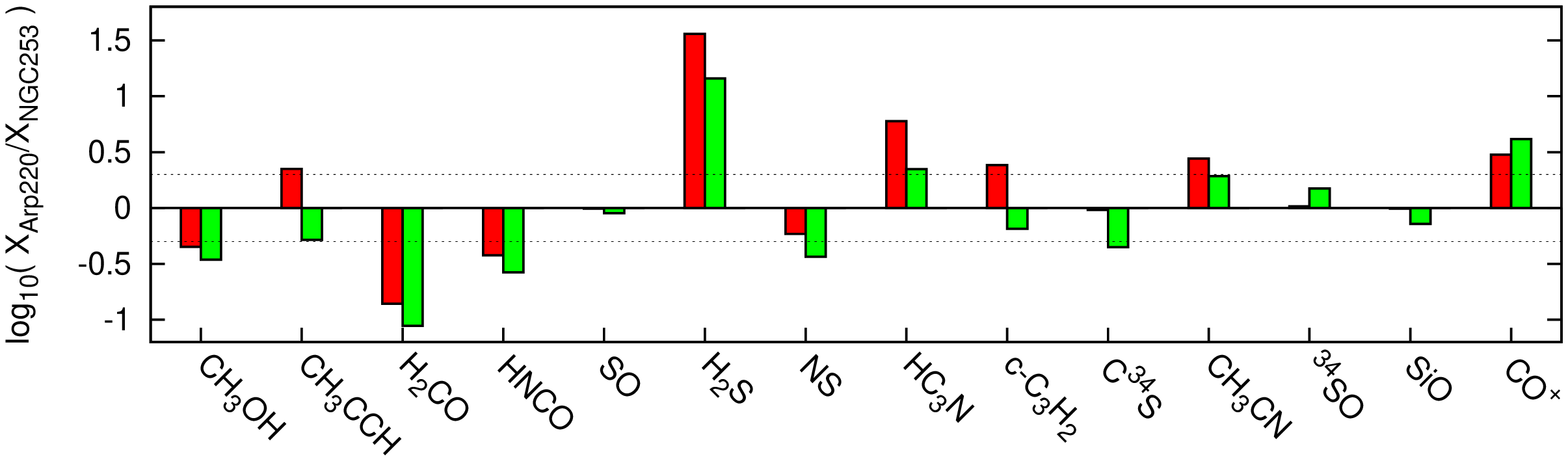}}
\caption{
({\it Upper panel}) Comparison of the fractional abundances relative to $^{13}$CS derived for the two velocity components in Arp\,220 and those
measured in the starburst galaxy NGC\,253.
({\it Lower panel}) Ratio of the relative abundances of each velocity component in Arp\,220 and in NGC\,253.
Horizontal dotted lines are set to indicate a ratio of factor of 2 with the aim of guiding the eye in the comparison. Differences below a factor of 2 are not
considered representative.
\label{fig.AbunCompCS}}
\end{figure}

In order to avoid the uncertainties derived from the estimate of the H$_2$ column density, we referred the relative abundances of the detected species
to that of CS.
The CS molecule traces the densest gas component within Arp\,220 \citep[$n\sim10^6\rm cm^{-3}$][]{Greve2009}, which is likely the source
of most of the observed chemical complexity.
The upper panel in Fig.~\ref{fig.AbunCompCS} shows a comparison of the abundances relative to the isotopologue $^{13}$CS in both Arp\,220 and NGC\,253.
The lower panel of Fig.~\ref{fig.AbunCompCS} shows the ratio between the abundances measured in Arp\,220 and those in NGC\,253.
In the comparison we assume that $^{13}$CS is only moderately optically thick ($\tau<1$) in both galaxies, and that both galaxies
have a similar $^{12}$C/$^{13}$C ratio.
Additional uncertainties might need to be considered in this comparison if the extent of the emission 
%varies or whether
or the considered excitation conditions 
%are not accurate 
vary among the different species.
Thus, only abundance differences above a factor of 2 will be considered as relevant between the two sources.
Such scatter in the chemical composition are found both in prototypical nearby galaxies and Galactic sources
as described in Sect. 4.4.1 and 4.4.2 in \citet{Mart'in2006}, respectively

As mentioned in Sect.~\ref{sect.CO}, we cannot spatially resolve the two nuclei.
However, we can infer some information from the double Gaussian fit to the line profile, 
given that both nuclei are somewhat separated in velocity.
We find that the abundances in both velocity components, relative to either H$_2$ or $^{13}$CS,  are the same within a factor of two.
But we also find an overabundance of a factor of $\sim 4$ in CH$_3$CCH and c-C$_3$H$_2$, and of a factor of $\sim 2.5$ in H$_2$S, and HC$_3$N
in the lower velocity component, presumably the western nucleus, 
% CH3CCH 4.3186200826151
% C3H2   3.7218168691966
% H2S    2.4990560684378
% HC3N   2.6788043777714

%When compared to NGC\,253 
We find that, with few exceptions, nearly all molecular species show similar abundances, within a factor of 2, to those in NGC\,253.
% between both objects.
% but for a few exception
The main differences, shown by both nuclei in Arp\,220, are the prominent overabundance of H$_2$S and the underabundance of H$_2$CO in Arp\,220 as
compared with NGC\,253.
To a lesser extent, we find the abundances of HC$_3$N (mostly in the velocity component 1) and CO$^+$ (for both velocity components) to be
a factor $\sim3$ larger in Arp\,220.
We also find a marginal underabundance of molecules such as CH$_3$OH and HNCO, being a factor $\sim2-3$ lower towards Arp\,220 than
%with respect to those measured 
in NGC\,253.
Overall we find that, with a few exceptions, the chemical composition of Arp\,220 bears a significant resemblance to that of NGC\,253.
%In the following discussion we aim to figure out whether the differences found are the evidence of an obscured AGN or the result of an extreme
%starburst within the central region of Arp\,220.

%\clearpage

\subsubsection{Sulfur chemistry in Arp\,220}
The molecular abundance that differs the most in Arp\,220 from
%as compared to 
the starburst galaxy NGC\,253 is that of H$_2$S.
We observed that all other
%the rest of 
sulfur bearing molecules in the survey, namely SO, NS, and the isotopologues C$^{34}$S and $^{34}$SO, appear to have
relative abundances similar to those in the nuclear region of NGC\,253.
Our observations show that the average relative abundance ratio is $\rm (H_2S/^{13}CS)_{\rm Arp\,220}\sim 20\times(H_2S/^{13}CS)_{\rm NGC\,253}$.
However this difference strongly depends on the excitation temperature of the species as their abundances have been derived from 2~mm transitions
in NGC\,253 and 1.3~mm for Arp\,220.
In order to evaluate whether the observed difference is not due to excitation and actually traces an abundance difference, we calculated the ratios
in both galaxies for different excitation temperatures of the gas.
Even in the extreme case where the H$_2$S would trace hot gas ($T_{\rm ex}>50$\,K), the ratio between the relative abundances in both galaxies
would drop down to a factor of $\sim 3-5$.
It shows that regardless of excitation considerations, the relative H$_2$S overabundance in Arp\,220 with respect to NGC\,253 would still be
significant.

Though absolute fractional abundances of the sulfur bearing species strongly depend on the accuracy of the H$_2$ column densities, the relative
abundances of these species can be reproduced by the available chemical models \citep{Hatchell1998}.
From the comparison to models, the overabundance of H$_2$S can have two possible explanations.
The first assumes the grain composition in Arp\,220 to be significantly different from that of NGC\,253. The H$_2$S frozen in grains, and subsequently
injected into gas phase, should be one order of magnitude more abundant.
The other possibility is 
%Otherwise, we can 
to assume the grain composition to be similar in both objects and consider time dependant sulfur bearing molecules abundances \citep{Hatchell1998}
to be tracing the average age of the star forming molecular clouds.
%as chemical clocks of hot core evolution.
Though it is somewhat speculative
%it would be uncertain 
to discuss time scales without an accurate measurement of the absolute abundances, we can try to explain the observed
differences in the framework of the evolutionary stage of the starburst.
In this scenario, the overabundance of H$_2$S would imply the hot cores in Arp\,220 are, on average,
in an earlier stage of evolution than NGC\,253.
The comparison of chemical models with observations of a small sample of massive dense cores by \citet{Herpin2009} shows how the CS/H$_2$S
abundance ratio decreases by an order of magnitude as a function of the stage of evolution of the clouds.
This result would also point out the H$_2$S difference observed between Arp\,220 and NGC\,253 as a consequence of evolutionary stage of the starburst.
From this observational result, the cores in Arp\,220, on average, would be in a later stage of evolution than NGC\,253,
contrary to our previous assertion. However, in that case we would not expect such high H$_2$S abundance as observed in Arp\,220.
Therefore, the large relative abundance of H$_2$S observed towards Arp\,220 appear to favour the idea of an earlier stage of starburst evolution
in this galaxy with respect to that in NGC\,253.
Nevertheless, the abundances of other sulfur-bearing species such as OCS, H$_2$CS and SO$_2$ would be needed for an accurate comparison with 
the chemical models \citep{Hatchell1998}, similar to that performed with NGC\,253 \citep{Mart'in2005}.

Regarding the possible AGN-driven origin of the enhanced H$_2$S abundance, there is no evidence in the literature to attribute the large
abundance of H$_2$S to a possibly hidden AGN.
Moreover, recent laboratory experiments show the rapid decline of H$_2$S observed in icy mantles when exposed to ion irradiation \citep{Garozzo2010}.

\subsection{Isotopic ratios in Arp\,220}
\label{sect.isotRat}
In Table~\ref{tab.IsotRatios} we summarize the isotopologue column density ratio derived for each of the species where any isotopic substitution has been measured.
Ratios are given for each of the velocity components as well as the average ratio. In the following we will only consider the average ratios, which are less
affected by the fitting constraints applied for modelling the observed spectrum.

\begin{table}
\begin{center}
\scriptsize
\caption{Isotopic column density ratios \label{tab.IsotRatios}}
\begin{tabular}{l c c}
\hline
\hline
Ratio                        & Vel. Components               &   Average           \\
\hline
$\rm ^{12}C^{16}O/^{13}CO$   &  $ 12.5   -    6.3  $         &  9.2                \\
$\rm ^{12}C^{16}O/C^{18}O$   &  $ 16.2   -    5.7  $         &  9.7                \\
$\rm HC_3N/H^{13}C_3N$       &  $  5.9   -    9.7  $         &  7.2                \\
$\rm CH_3CN/^{13}CH_3CN$     &  $ >15.0  -  >15.1  $         &  $>15.1$            \\
$\rm C^{34}S/^{13}CS$        &  $  3.3   -    1.4  $         &  2.0                \\
$\rm SO/^{34}SO $            &  $  4.6   -    2.8  $         &  3.2                \\
$\rm SiO/^{29}SiO$           &  $  2.8   -    3.2  $         &  3.0                \\
\hline
\end{tabular}
\end{center}
\end{table}

The large $\rm ^{12}CO/^{13}CO$ ratio ($R_{12/13}>20$) observed in luminous mergers as compared to normal starbursts has lead to the postulation of an intrinsic
differenciation between the ISM properties in these objects. %extreme starbursts.
An abundance difference can be understood if the ISM in mergers is fed with unprocessed molecular gas drawn from the external region
into their nuclear region, or the ISM is enriched in $^{12}$C by nucleosynthesis in newly formed massive stars \citep{Casoli1992}.
Additionally, these high ratios can otherwise be explained by different gas surface densities resulting in differences in opacities between luminous mergers
and starbursts galaxies \citep{Aalto1995}.
%However, our results on Arp\,220 shed some doubt on the ratios measured towards this galaxy.
Towards Arp\,220, \citet{Casoli1992} measured a $R^{2-1}_{12/13}$ ratio of $18\pm5$ based on the $J=2-1$ transition, consistent with the limit of
$R^{1-0}_{12/13}>19$ and $>22$ ratios based on the  $J=1-0$ lines \citep{Aalto1991,Casoli1992}.
Our results give a ratio $R^{2-1}_{12/13}\sim9$ towards Arp\,220.
%, which is a factor of two lower than the average value from the literature of $18\pm3$ \citep{Greve2009}.
However, this is based on the extracted spectra at the peak of emission, and an estimate of the extended emission is required to compare with the derived
single-dish ratio.
As detailed in Sect.~\ref{sect.CO}, the overall extended CO emission is about a factor of 2 larger than that measured towards the central position.
To estimate the extended $^{13}$CO emission, it would require an accurate continuum measurement and subtraction as well as precise determination
of the emission of the other molecules contributing to this crowded spectral band.
Thus, with the $^{13}$CO value at the central peak of emission and the integrated extended emission of CO we can constrain the ratio
as $9\leq R^{2-1}_{12/13}\leq 18$.
These limits assume the extreme situations of $^{13}$CO being as extended as CO or compact, respectively.
Though the upper limit agrees with 
%is still consistent with 
the average value derived from single dish in the literature
of $18\pm3$ \citep{Greve2009}, %the ratio might fall in 
our limits are consistent with the 
%in the upper 
range of the $10\la R^{1-0}_{12/13}\la 15$ ratio found in normal starbursts by \citet{Aalto1995}.
The $^{13}$CO emission from single dish observations may have been underestimated by both the baseline subtraction uncertainty
due to the limited spectral bandwidth and the unknown contamination by other species,
as well as to a relative calibration issues affecting the single dish data.
%However, large ratios such as that of $R^{2-1}_{12/13}\sim30$ towards NGC\,3256 \citep{Sakamoto2006a} still cannot be explained by these uncertainties.
%cannot be explained by the uncertainty in the continuum determination.

%However, other observed molecular species also point out towards a low carbon isotopic ratio.
The $\rm H^{12}C_3N/H^{13}C_3N$ ratio is $\sim7$, even lower than that measured with CO.
With a critical density of $n_{\rm crit}\sim2-5\times10^{6}\rm cm^{-3}$, the HC$_3$N emission is likely tracing the densest molecular material in Arp\,220.
Assuming both CO and HC$_3$N isotopologue ratios are linked to the carbon isotopic ratio, the similar measured ratio would imply equally large opacities
affecting the main isotopologues of both species.
Thus, no reliable estimates of the $^{12}$C/$^{13}$C isotopic ratio can be derived from these species.

If we assume the SO/$^{34}$SO ratio of $\sim3$ is less affected by opacity and is representative of the sulfur $\rm ^{32}S/^{34}S$, it would result in an extremely low
sulfur isotopic ratio, significantly lower than the ratios of $8-13$ measured in the starbursts galaxies NGC\,253 and NGC\,4945 \citep{Wang2004,Mart'in2005}.
This low ratio would be consistent with the scenario of $^{34}$S overproduction in supernovae \citep{Chin1996}.
Indeed, the supernova rate estimated towards Arp\,220 of $4\pm2\rm\,yr^{-1}$ \citep{Lonsdale2006}
is one or two orders of magnitude higher than the rates of $0.05-0.3\rm\,yr^{-1}$ towards NGC\,253 \citep{Ulvestad1997,Mattila2001}.
However, if we assume this $\rm ^{32}S/^{34}S$ ratio and the measured average ratio of $\rm C^{34}S/^{13}CS\sim2$, we estimate
a $\rm CS/^{13}CS$ ratio of $\sim6$, similar to the values derived from both CO and HC$_3$N.
Therefore we suspect $\rm C^{34}S$ and SO to be also affected by significant optical thickness effects, resulting in a lower limit to the ratio of SO/$^{34}$SO$>3$.

We measure a $\rm ^{13}CO/C^{18}O$ ratio of $\sim 1$, in agreement with the ratio derived with the $J=1-0$ and $J=2-1$ transitions \citep{Greve2009,Matsushita2009}.
%Since we do not have an accurate estimate of the opacity of the $^{13}$CO transition, we can only derive lower limits to the oxygen isotopic ratio.
Even if we assume an optical depth of $\rm ^{13}CO$ of $\tau_{13}>1$ \citep{Greve2009} and $^{12}$C/$^{13}$C isotopic ratio of $\sim40-50$, as measured
in starburst galaxies \citep{Henkel1993,Henkel1993a}, we infer a limit to the ratio of $\rm ^{16}O/^{18}O>80-100$.
This ratio is lower than that derived towards the starburst NGC\,253 \citep[$\rm ^{16}O/^{18}O>150$,][]{Harrison1999}, which might imply a large
opacity affecting $\rm ^{13}CO$ and even a significant opacity affecting $\rm C^{18}O$ as suggested by \citet{Matsushita2009}.
%which would again suggest an extreme ISM enrichment \citep{Henkel1993a} as a consequence of the intense starbursts in Arp\,220.

Additionally, we have also detected $^{29}$SiO at a very low ratio with respect to the main isotopologue.
Though the detection remains tentative, we can evaluate the feasibility of this detection as compared with the ratios measured in Galactic sources.
\citet{Wolff1980} found $\rm^{28}Si/^{29}Si=10\pm3$ toward bright SiO emission regions, namely Sgr~A and the massive star forming regions Sgr~B, Orion A, and W51.
They observed this ratio to be a factor of 2 below the terrestrial abundance.
Other studies find integrated intensity ratios of $5-8$ and 12 towards the Giant Molecular Clouds cores W3(H$_2$O) and Sgr~B2(M) \citep{Helmich1997,Nummelin2000},
and $10-20$ towards the late carbon star IRC~+1026 \citep{Groesbeck1994,Kawaguchi1995}.
Our detection results in a ratio $\rm^{28}Si/^{29}Si\sim3$ which is only a factor of 2 below the ratio
measured towards Orion~A \citep{Wolff1980}.
Even though this result has to be confirmed at other frequencies and the ratio is subject to a significant error due to the low signal-to-noise of the
lines involved, the measured ratio would tentatively point towards a slight $^{29}$Si enrichment in Arp\,220.
However this result is tentative.

As shown for NGC\,253 by \citet{Mart'in2010a}, an accurate measurement of isotopic ratios towards Arp\,220 will require the observation of doubly substituted isotopologues.
Our results suggest that the abundances of $\rm^{34}S$ and $\rm^{18}O$ isotope could be similar or even higher than those found in local starbursts.
A sequence of short and intense starbursts \citep{Parra2007} or previously enriched gas feeding the starburst might be the reason for such enriched ISM towards Arp\,220.
However the opacity of the main isotopologues prevents us from further constraining these isotopic ratios.

\subsubsection{H$_2^{18}$O: water vapor isotopologue emission}
\label{sect.watermaser}
The spectral feature observed at 203.4\,GHz has been identified as the $^{18}$O isotopologue of the water emission reported by \citet{Cernicharo2006}.
Table~\ref{tab.Water} shows the parameters of the Gaussian profiles fitted to both H$_2$O and H$_2^{18}$O.
The emission of para-H$_2^{18}$O  $3_{1,3}-2_{2,0}$ has been detected in a number of Galactic hot cores where the emission is observed to be significantly
blended to the brighter SO$_2$ $12_{0,12}-11_{1,11}$ emission $\sim 16\rm\,MHz$ below ($+24\rm\,km\,s^{-1}$) the H$_2^{18}$O line \citep{Jacq1988,Gensheimer1996}.
We do not find any trace of significant SO$_2$ emission in the frequency range covered, where a number of SO$_2$ transition, brighter than that at 203\,GHz,
should have been detected. Moreover, if the observed feature would be mostly dominated by SO$_2$ it would appear red-shifted with respect to the H$_2$O velocity.
However, the rarer isotopologue emission is slightly narrower and blue-shifted ($\sim -65\rm\,km\,s^{-1}$) with respect to the H$_2$O.
This is consistent with the observed H$_2$O profile \citep{Cernicharo2006}, which is brighter at the lower velocities.
Also the contribution from methyl formate and dimethyl ether, significant in Galactic hot cores \citep{Gensheimer1996},
is negligible in Arp\,220 based on the fitting of these species to the whole 40\,GHz band.

\begin{table}
\begin{center}
\scriptsize
\caption{Water vapor isotopologues detections \label{tab.Water}}
\begin{tabular}{l @{\quad} c @{\quad} c @{\quad} c @{\quad} c @{\quad} c @{\quad} c}
\hline
\hline
Isotopologue                                 & $\nu$   &  $\int S_\nu\,{\rm d}v$  &    $V_{\rm LSR}$             &  $\Delta v_{1/2}$   & $S_\nu$     \\
                                             & GHz     &  Jy\,km\,s$^{-1}$        &    km\,s$^{-1}$              &  km\,s$^{-1}$       &  mJy         \\
\hline
$\rm H_2O$       $3_{1,3}-2_{2,0}$ \tablefootmark{a} & 183.310 &   56.1                   &    $\sim5400$                &   310               &  170       \\
$\rm H_2^{18}O$  $3_{1,3}-2_{2,0}$                   & 203.407 &    9.5                   &     5335                     &   246               &  36        \\
\hline
\end{tabular}
\tablefoot{
\tablefoottext{a}{From \citet{Cernicharo2006}}
}
\end{center}
\end{table}

Unlike nearby starburst galaxies, where no H$_2^{18}$O observations have been reported,
the detection of H$_2^{18}$O emission towards Arp\,220 is only possible due to the 
heavily enriched molecular gas in this galaxy, with very low $\rm^{16}O/^{18}O$ as suggested in Sect.~\ref{sect.isotRat}.
We derive an integrated flux ratio of $\rm H_2O/H^{18}_2O\sim 6$. %, which is within a factor of two of the estimated limit to the oxygen isotopic ratio.
This ratio is not representative of the $^{16}$O/$^{18}$O isotopic ratio given the 
%optically thick 
weak maser origin of the H$_2$O line.
Moreover, the ratio between both isotopologues might imply that most of the observed $\rm H_2O$ emission in Arp\,220 is 
%thermal and not 
indeed a weak maser.
%as claimed by \citet{Cernicharo2006}.

From the $\rm H_2^{18}O$ integrated intensity of $\sim70.5\rm\,K\,km\,s^{-1}$ measured in this work we can estimate a total column density
$N_{\rm H_2^{18}O}=4.6\times10^{16}\rm\,cm^{-2}$ assuming optically thin emission and an excitation temperature of $T_{\rm ex}=100-200$\,K.
Given that we use the C$^{18}$O emission as a tracer of the total H$_2$ column density, we can reliably derive the fractional abundance of water of
$X_{\rm H_2O}=1.7\times10^{-5}$, independent of the assumption on the $^{16}$O/$^{18}$O ratio.
This value agrees well with the fractional abundance observed towards the Sgr~B2 region \citep{Cernicharo2006b} and with an average of $\sim 10^{-5}$
towards the 13 hot cores where $\rm H_2^{18}O$ was detected in the sample of \citet{Gensheimer1996}.

% 9.5 Jy km/s = 70.49 K km/s
% 2''=698 pc @ 72 Mpc
% Area = 698^2*1.133
% Total = 70.49*698^2*1.133=
We observed a $\rm H_2^{18}O$ luminosity of
%$9.7\times10^6\rm\,K\,km^{-1}\,pc^2$ 
$3.9\times10^7\rm\,K\,km^{-1}\,pc^2$ in the central $2''$ ($\sim 700$\,pc) of Arp\,220.
The observations of $\rm H_2^{18}O$ by \citet{Gensheimer1996} towards the Sgr~B2 molecular cloud complex
shows the emission to be concentrated towards Sgr~B2(N) hot core, with a significantly lower intensity towards the Sgr~B2(M) core.
They reported an integrated intensity of $17.4\rm\,K\,km\,s^{-1}$ towards Sgr~B2(N).
The emission is unresolved at their 12$''$ ($\sim 0.5\,pc$) resolution, which is similar to the $0.4-0.5$\,pc size estimated for the northern
core \citep{Lis1990}.
Assuming a range in the extent of the emission extent of $0.5-1$\,pc and using a similar argument to that for H$_2$O in \citet{Cernicharo2006},
a total of $\sim2-8\times10^6$ Sgr~B2(N)-like cores would be required and enclosed in the central 700\,pc of Arp\,220 to explain the observed $\rm H_2^{18}O$ emission.
If we consider the luminosity of Sgr~B2(N) of $\sim 10^7\,L_\odot$ \citep{deVicente2000}, such number of massive star forming cores alone would easily account for the bolometric
luminosity of Arp\,220.

%Regarding the compactness of the star forming cores in Arp\,220,
Such a concentration of hot cores would require an average distance of $3.5-5.6$\,pc between cores, assuming they
are distributed isotropically within a spherical region of 700~pc diameter.
Assuming them to be distributed within a disk with a thickness of $\sim200$~pc would yield an average
distance of $2.4-3.7$~pc.
These rough estimates are still larger than the projected distance of $\sim 1.6$~pc between the hot cores in the 
Sgr~B2 molecular complex within the Galactic center region.
Therefore the estimate of the number of hot cores in Arp\,220 does not put severe constraints on the
compactness of the star forming regions in this galaxy.
However, it does imply a widespread intense star formation throughout its whole central region.

%To our best knowledge the H$_2^{18}$O has never been identified in the ISM so no direct flux comparison with Galactic sources is possible.
%The very low resolution ($\sim300\rm\,km\,s^{-1}$) spectral line survey by \citep{Serabyn95} toward Orion~IRc2 covered
%the H$_2^{18}$O frequency.
%However the resolution and line confusion is not enough to identify the $\sim 40\rm\,km\,s^{-1}$ water feature \citep{Cerni1994,Cerni1999}.

%If we assume the lower limit ratio $\rm ^{16}O/^{18}O>12$ (see Sect.~\ref{sect.isotRat}), our measurment would imply 
%an integrated water line emission of $\geq850\rm\,K\,km\,s^{-1}$ in the central $2''$ ($\sim 700$\,pc) of Arp\,220.
%Using a similar argument to that found in \citet{Cernicharo2006}, this means that the central region of Arp\,220 houses
%between $4\times10^5$ and $>4\times10^6$ Sgr~B2 like hot cores.
%This constraint is a factor $>2$ than that derived from the main isotopologue.

\subsection{High temperature vibrationally excited molecular gas}\
\label{sect.highTemp}
Vibrationally excited HC$_3$N emission probing the hot molecular component has been detected, so far, towards two extragalactic sources, 
the luminous infrared galaxy (LIRG) NGC\,4418 and the ULIRG Arp\,220 \citep[][Mart\'in-Pintado, in prep]{Costagliola2010}.
Both galaxies show prominent silicate absorption features \citep{Roche1986,Smith1989} which can be a sign of high temperature dust heated by hot young stars \citep{Spoon2006}.
Within the Galaxy, vibrationally excited HC$_3$N emission is found towards massive \citep[Sgr~B2 and Orion][]{deVicente2000,deVicente2002} and intermediate \citep[HW2,][]{Mart'in-Pintado2005}
star forming regions.
Towards Arp\,220 we find HC$_3$N and CH$_3$CN probing hot molecular gas with derived temperatures of $T_{\rm ex}=300-400$\,K.
The critical densities of the HC$_3$N observed transitions are $n_{\rm crit}>10^6\rm cm^{-3}$ \citep{Wernli2007}.
Thus, these species are tracing the densest and hottest molecular gas in Arp\,220.

In Table~\ref{tab.NTrot} two sets of physical parameters are derived for HC$_3$N.
The first one takes only into account the pure rotational transitions of HC$_3$N for which we derive a rotational temperature of $T_{\rm rot}\sim 35$\,K.
The second set of parameters with a temperature of  $T_{\rm rot}\sim 350$\,K are derived with the vibrationally excited transitions only.
As observed towards the Sgr~B2 molecular cloud in the Galactic Center region, both HC$_3$N and CH$_3$CN rotational emission arise from the same region and trace a molecular
ridge containing the massive star forming regions \citep{deVicente1997}.
Vibrational emission, on the other hand, mostly arises from the hot cores, though it is not restricted
to the core extent \citep{deVicente2000}.
In order to populate the vibrationally excited levels of the observed transitions,
HC$_3$N and CH$_3$CN require IR photons in the range of $15-45~\mu m$ and $20-30~\mu m$, respectively.
While the rotational transitions arise from the warm envelope around the hot cores \citep[$40-80$\,K, $\rm 2\times10^5\,cm^{-3}$;][]{deVicente1997},
the hot component will be directly tracing the innermost regions around the hot cores with temperatures well above 200\,K \citep{deVicente2000}.
Thus, the excitation of both HC$_3$N and CH$_3$CN in Arp\,220 agrees with its emission originating in Sgr~B2-like regions.

\section{Conclusions: AGN vs SB driven chemistry}

%This detection turn $\rm H_2^{18}O$ into an ideal tracer of the water distribution and kinematics in Arp\,220, being at a frequen.
Up to now, the only available extragalactic molecular studies available as comparison templates are the unbiased study toward NGC\,253 \citep{Mart'in2006}
and M\,82 (Aladro et al., in prep), and the targeted observations towards NGC\,4945 \citep{Wang2004}.
These studies have 
%all being aimed toward 
characterized the chemical complexity of starburst galaxies.
The molecular abundances in Arp\,220, resulting from the detected transitions in the 1.3~mm atmospheric window, do resemble those found towards the
starburst galaxy NGC\,253.
The only species showing an outstanding abundance towards Arp\,220 is H$_2$S.
Although the abundance of H$_2$S has only been measured towards NGC\,253, this overabundance is likely to be the result of grain disruption
and subsequent H$_2$S injection into gas phase in the early stages of star formation \citep{Hatchell1998}. This sets the burst of 
star formation in Arp\,220 in a very early stage of evolution and thus, on average, younger than the starburst in NGC\,253.
This comparison suggests that the chemistry in Arp\,220 is mostly driven by bursts of star formation
where differences in the observed abundances
%are similar to those observed between different starbursts and 
can
%therefore 
be accounted for by a difference in the state of evolution of the starbursts \citep{Mart'in2006}.  

Unfortunately, no complete chemical templates of the ISM in the vicinity of AGNs are available and therefore the best tracers in highly X-ray irradiated regions
are based on chemical models.
The results from such models \citep{Meijerink2005,Meijerink2006,Meijerink2007,Loenen2008} have motivated a number of observational
studies focused on the species CO, HCN, HNC, HCO$^+$, and CN \citep{P'erez-Beaupuits2007,Krips2008,Baan2008,P'erez-Beaupuits2009}.
The HCN/HCO$^+$ ratio was used by \citep{Aalto2007a} to evaluate the presence of an AGN on the LIRG NGC\,4418 which, like Arp\,220,
has a heavily obscured nucleus.
The HCN and HCO$^+$ comparative study by \citet{Krips2008} shows how the HCN/HCO$^+$ ratios in Arp\,220 are similar
to those measured in NGC\,1068, the prototypical AGN galaxy. However the excitation of HCN differs significantly
in both galaxies \citep{Krips2008}.
On the other hand, though overluminous HNC emission in Arp\,220 could be attributed to either pumping through mid-IR
photons or the presence of an XDR region \citep{Aalto2007}, the observed HNC/HCN ratio suggest a PDR rather than a XDR origin \citep{Baan2008,Baan2010}.
% HCN Krips 2008 = 1.8E15 +-0.1E15
%Within our survey we observed the emission of CN.
We estimate a column density ratio CN/HCN$\sim0.5$, where we have used the single-dish HCN observations by \citet{Krips2008}
and our survey measurement of the CN emission.
To estimate the HCN column density we assumed a source size of $2''$ similar to the one used in this paper.
This ratio is lower than the range CN/HCN$\sim1-4$ that \citep{P'erez-Beaupuits2009} claims as indicative of
XDR regions.
Therefore, the CN emission does not appear to be enhanced as observed in the circumnuclear disk of NGC\,1068 \citep{Garc'ia-Burillo2010}.
Moreover, \citet{Garc'ia-Burillo2010} found an enhancement of SiO in this region, which does not seem to be particularly prominent in Arp\,220
as compared to NGC\,253.
%,  which could lead to the presence of an additionaly heating mechanism.
Therefore, we do not find any molecular emission that could be attributed to X-ray driven chemistry in the nuclear
region of Arp\,220.

Measured isotopologues ratios appear to point to both large opacities affecting the main isotopologues of most observed species and
an enriched molecular material in the nuclear region of Arp\,220.
ISM enrichment towards Arp\,220 is similar or even enhanced with respect to that found in other starbursts, such as NGC\,253 and M\,82,
as a consequence of a series of consecutive short and intense starburst events \citep{Parra2007}.
Among the isotopologues, the observed H$^{18}_2$O luminosity can be accounted for by the emission of a few $10^6$ hot molecular
cores associated with the massive star formation within the central 700 pc of Arp\,220.
Such a concentration of cores could be responsible for the whole bolometric luminosity of this galaxy, 
rendering unnecessary a significant contribution to the luminosity by a deeply embedded AGN.
%would be negligible.
Far from the H$_2$O atmospheric absorption, the optically thin emission of H$^{18}_2$O is proven to be one of the best tracers
of massive star forming hot cores in highly obscured nuclei of galaxies.

The importance of the star formation within the nuclear region of Arp\,220
is further supported by the detection of vibrationally excited emission of HC$_3$N and CH$_3$CN, with vibrational
temperatures $>300$\,K.
Such emission is also tracing the molecular component associated with hot cores as reported by Mart\'in-Pintado (in prep.).
However, vibrationally excited HC$_3$N emission with a temperature of 500~K towards NGC\,4418 could be understood
as a compact and deeply embedded AGN, heating up the surrounding material \citep{Costagliola2010}.
It is remarkable that such vibrationally excited emission has never been reported towards nearby starbursts.
The detection towards the ULIRG Arp\,220 and the LIRG NGC\,4418, might be due to the significantly larger contribution
of hot core emission as a consequence of the higher star formation rates in these galaxies. Large opacities affect the pure
rotational transitions of HC$_3$N (Sect.~\ref{sect.isotRat}).
Due to the large continuum opacity at 1mm \citep{Downes2007}, the molecular emission we observed cannot arise from the vicinity of an AGN.
Indeed, the simple models presented \citet{Schleicher2010} show the heating sphere of influence of a supermassive AGN to be limited to the central
$\sim100$\,pc while the starburst heating dominates outside this volume.
Therefore, the hot gas where the vibrationally excited emission of CH$_3$CN and HC$_3$N cannot be tracing the regions around the Compton thick AGN
but must arise from other regions likely unaffected by the AGN radiation and purely heated by the starbursting events in Arp\,220.

\section{Summary}
In this paper we present the study of a 1.3\,mm wavelength scan of Arp\,220 covering the 40\,GHz range between the rest frequencies of 202 and 242\,GHz.

Molecular line emission is detected over more than 80\% of the observed band.
Continuum emission, previously overstimated in several works due to line contamination, has been estimated from
the few line free regions of the spectrum.
We derive an average 1.3~mm continuum flux of 142~mJy, ranging from 129 to 154~mJy across the band.
The total line contribution to the overall flux measured in the 40~GHz band is estimated to be $\sim28\%$, while
CO emission alone accounts for only $\sim9\%$.
Therefore line contamination to thermal continuum is shown to be more critical than the non-thermal contribution which
only counts for a $3-6\%$ of the total flux density at these frequencies.

We present maps of the CO emission detected in the $V_{\rm LSR}$ range between 4950 and 5750 \kms.
This map likely recovers all the single-dish detected flux thanks to the short baselines in the SMA sub-compact configuration.

A total of 73 spectral features are detected, implying a detection rate of 1.8 lines/GHz.
15 molecular species and 6 isotopologues are identified.
The $^{13}$C isotopologues of HC$_3$N, as well as H$_2^{18}$O, $^{29}$SiO, and CH$_2$CO, are reported for the first time in
the extragalactic ISM.
The lines in the entire 40~GHz spectrum have been fitted to the molecular emission of the identified species assuming LTE conditions.
Of the $\sim3000$ transitions used to perform the fit, only 163 are detected above a $1\,\sigma$ level.

The molecular composition and abundances derived from the modelling favor a purely starburst 
%activity ocurring 
powering source in Arp\,220.
This conclusion is based on the direct comparison with the molecular abundances derived for the nearby starburst galaxy NGC\,253,
as well as the detection of the water isotopologue, H$^{18}_2$O, and vibrationally excited emission of HC$_3$N and CH$_3$CN, which are likely
tracing the hot cores associated with the massive star formation sites within the nuclear region of Arp\,220.
The comparison with NGC\,253 chemistry shows a prominent overabundance of H$_2$S which can be plausibly interpreted as Arp\,220 being in a very early
stage of its burst of star formation, as predicted by chemical models \citep{Hatchell1998}.

Though deep sensitive observations are required to accurately constrain the isotopic ratios in Arp\,220, our measurements suggest
the presence of an isotopic enrichment similar to or more prominent than that towards nearby starbursts.
The large opacities of the main isotopologues prevent us from
getting further constraints from the observed isotopologue abundance ratios. 
Indeed an effect of the opacity in HC$_3$N likely similar to that in CO is found.

From our H$^{18}_2$O and C$^{18}$O measurements we estimate a water abundance of $X_{\rm H_2O}=1.7\times10^{-5}$ without assumptions
on the $^{16}$O/$^{18}$O isotopic ratio. This abundance agrees with that derived towards the Galactic hot cores
\citep{Gensheimer1996,Cernicharo2006b}.
Assuming all the H$^{18}_2$O emission comes from the massive star forming hot cores within the nuclear region of Arp\,220, the
H$^{18}_2$O luminosity of $3.9\times10^7\rm\,K\,km^{-1}\,pc^2$ within the central $2''$ would require a $\sim2-8\times10^6$ Sgr~B2-like
hot cores enclosed within its inner 700~pc.
The detected vibrationally excited transitions of HC$_3$N and CH$_3$CN trace a molecular component with temperatures between
300 and 400~K. Within the Galaxy, such emission is only observed from hot molecular cores \citep{deVicente2000,deVicente2002}.

We have not found chemical evidence for the presence of a buried AGN in the nuclei of Arp\,220.
Though likely present as suggested in various studies \citep[][and references therein]{Downes2007},
the suspected black hole has no observable effects in the heating of the nuclear ISM in this galaxy.
Moreover, the starburst contribution alone would be able to explain the huge infrared emission observed in Arp\,220.

\begin{acknowledgements}
The Submillimeter Array is a joint project between the Smithsonian Astrophysical Observatory and the Academia Sinica Institute of Astronomy 
and Astrophysics and is funded by the Smithsonian Institution and the Academia Sinica.
\end{acknowledgements}

%{\it Facilities:} \facility{SMA ()}

%\appendix

%\section{Appendix material}

\bibliographystyle{aa}	% see astronat package, apj.bst
\bibliography{ms.bib}	% links to ngc6000.bib for bibtex information

\end{document}